\documentclass[12pt,english]{article}
\usepackage[english]{babel}
\usepackage{amsmath,amssymb,amscd}
\usepackage{amsfonts}
\usepackage{graphicx}
\usepackage{appendix}
\usepackage{url}
\usepackage{xspace}
\usepackage{cite}
\makeatletter
\renewcommand\section{\@startsection {section}{1}{\z@}%
                                   {-5.5ex \@plus -1ex \@minus -.2ex}
                                   {2.3ex \@plus.2ex}%
                                   {\normalfont\large\bfseries}}
\renewcommand\subsection{\@startsection{subsection}{2}{\z@}%
                                     {-3.25ex\@plus -1ex \@minus -.2ex}%
                                     {1.5ex \@plus .2ex}%
                                     {\normalfont\bfseries}}

\numberwithin{equation}{section}
\makeatother
\date{}
\newcommand{\ba}{\begin{align}}

\newcommand{\bea}{\begin{eqnarray}}
\newcommand{\eea}{\end{eqnarray}}
\newcommand{\be}{\begin{equation}}
\newcommand{\ee}{\end{equation}}
\newcommand{\eq}[1]{(\ref{#1})}

\newcommand{\C}{{\mathbb C}}

\newcommand{\ie}{{\it i.e.~}}
\newcommand{\eg}{{\it e.g.~}}
\def\eq{\begin{equation}}
\def\en{\end{equation}}
\def\eqa{\begin{eqnarray}}
\def\ena{\end{eqnarray}}
\def\aeq#1{\begin{align}#1\end{align}}  
\def\ateq#1#2{\begin{alignat}{#1}#2\end{alignat}}  
\def\ignorethis#1{}

\def\bra#1{\langle #1 |}
\def\ket#1{| #1\rangle}
\def\expval#1{\langle \, #1 \,\rangle}

\def\Complexes{\mathbb{C}}		
\def\Reals{\mathbb{R}}			
\def\Integers{\mathbb{Z}}		
\renewcommand{\Im}{\mathrm{Im}} 
\DeclareMathOperator{\Ker}{Ker}	

\def\Aone{\textbf{A1}\xspace}
\def\Atwo{\textbf{A2}\xspace}

\def\COne{\textbf{C1}\xspace}
\def\CTwo{\textbf{C2}\xspace}

\def\so{\mathbf{so}}
\def\u{\mathbf{u}}
\def\sl2{\mathbf{sl}(2)}
\DeclareMathOperator{\sgn}{sgn}

\makeatletter
\newcommand{\overleftrightsmallarrow}{\mathpalette{\overarrowsmall@\leftrightarrowfill@}}
\newcommand{\overrightsmallarrow}{\mathpalette{\overarrowsmall@\rightarrowfill@}}
\newcommand{\overleftsmallarrow}{\mathpalette{\overarrowsmall@\leftarrowfill@}}
\newcommand{\overarrowsmall@}[3]{%
  \vbox{%
    \ialign{%
      ##\crcr
      #1{\smaller@style{#2}}\crcr
      \noalign{\nointerlineskip}%
      $\m@th\hfil#2#3\hfil$\crcr
    }%
  }%
}
\def\smaller@style#1{%
  \ifx#1\displaystyle\scriptstyle\else
    \ifx#1\textstyle\scriptstyle\else
      \scriptscriptstyle
    \fi
  \fi
}
\makeatother

\title{\vspace{-1cm}\begin{flushright}{\small RUNHETC-2015-10}\end{flushright}\vspace{2cm}
\LARGE Cauchy conformal fields in dimensions $d>2$
}

\author
{
Daniel Friedan$^{1,2}$\footnote{friedan@physics.rutgers.edu}\ \ and  Christoph A.~Keller$^{1,3}$\footnote{christoph.keller@math.ethz.ch}
\\
\\
$^1$NHETC and Department of Physics and Astronomy\\
Rutgers, The State University of New Jersey\\
Piscataway, New Jersey 08854-8019, USA\\
\\
$^2$The Science Institute, The University of Iceland,
Reykjavik, Iceland\\
\\
$^3$Department of Mathematics, ETH Zurich, 8092 Zurich, Switzerland
}

\begin{document}

\maketitle

\begin{center}
{\bf Abstract}
\end{center}
Holomorphic fields play an important
role
in 2d conformal field theory.
We generalize them to $d>2$
by introducing the notion of Cauchy conformal fields,
which satisfy a first order differential
equation such that they are determined everywhere once
we know their value on a codimension 1 surface.
We classify all the unitary Cauchy fields. 
By analyzing the mode expansion on the unit sphere,
we show that all unitary Cauchy fields are free
in the sense that their correlation functions
factorize on the 2-point function.
We also discuss the possibility 
of non-unitary Cauchy fields and classify them
in $d=3$ and 4.

\newpage
\tableofcontents
\newpage

\section{Introduction}
\subsection{Generalizing holomorphic fields to $d>2$ -- the Cauchy condition}
Holomorphic quantum fields
play a major role
in two dimensional conformal field theory.
Their defining property is the Cauchy-Riemann equation 
\be\label{Cauchy2d} 
(\partial_{x} + i
\partial_{y})\phi(z)=0\,.
\ee
Equation (\ref{Cauchy2d}) defines a Cauchy problem.  Namely, if we
know the field $\phi$ on some contour, say for instance $x=\mathit{const}$, or
the circle $|x+iy| = \mathit{const}$, then (\ref{Cauchy2d}) uniquely determines $\phi$
everywhere.
That $\phi$ depends holomorphically on the single complex variable $z=x+iy$ 
is equivalent to the Cauchy-Riemann equation (\ref{Cauchy2d}).
A 2d holomorphic field has a mode expansion
\eq
\phi_{n} = \int_{C} dz \; f_{n}(z) \phi(z)
\,,\qquad
\phi(z) = \sum_{n}  f^{*}_{n}(z) \phi_{n}
\,,\qquad
\int_{C}  f_{m}(z) f^{*}_{n}(z) = \delta_{m,n}
\en
where the smearing functions $f_{n}(z)$ are a complete set of 
functions holomorphic in some neighborhood of the contour $C$ (for 
example, a circle in the 2d space-time).
The Cauchy property allows us to 
deform the contour without changing the result, within the region 
where $\phi(z)$ and $f_{n}(z)$  are non-singular.
By appropriately deforming the contour, we can thus
obtain the commutation relations of the modes
from just the singular terms in the operator products of the 
holomorphic fields.
The representation theory of the algebra of the modes becomes a powerful tool for analyzing the 
quantum field theory.
In a complementary vein, the global properties of the Cauchy-Riemann equation 
put stringent constraints on the correlation functions of holomorphic 
fields, making possible exact solutions.

Holomorphic quantum fields were first constructed and 
studied in string 
theory
\cite{Fubini:1969wp,Fubini:1971ce,DelGiudice:1971fp,Brower:1972uf}.
Early uses of contour deformation techniques can be found in
\cite{Corrigan:1973jz,Brink:1973nb}.
P.~Goddard wrote
``People
gradually realized through this time (1971-73) that one could more
profitably use the analytic properties of the fields in this way than
think in terms of distributions on the unit circle'' 
\cite{GoddardPrivate}.
Holomorphic fields and the contour deformation technique were 
later rediscovered in \cite{Friedan1982c}.

In this paper our goal is to generalize the notion of holomorphic conformal fields to higher
dimensions, and to attempt a classification of such fields.  
Our hope -- unrealized -- was that there might be as rich a variety of such 
fields  as in two dimensions.
To mimic the two dimensional case,
we search for fields $\phi$ that satisfy a
first order differential equation which has the Cauchy property ---
the property of uniquely determining $\phi$ everywhere once
$\phi$ is known on some codimension 1 surface $S$.
This Cauchy property generalizes to $d>2$,
along with the techniques of mode expansion and contour deformation.
The property of depending on a single complex variable on the other hand
does not seem to generalize to conformal fields in $d>2$.

Not every field satisfying a first order differential
has the Cauchy property. 
For example, a conserved spin 1 current $j^\mu(x)$ satisfies
the first order equation
$\partial_\mu j^\mu = 0$,
but knowing $j^{\mu}$ on the hyperplane $x^d=0$ is not enough to determine 
$j^{\mu}$ on nearby hyperplanes.
We are thus looking for fields which satisfy a special
type of first order differential equation.

An example of a field that does have the Cauchy
property is
a self-dual two form $\phi^{\mu\nu}(x)$
in $d=4$ dimensions.
It satisfies
the self-duality condition
\eq
\phi^{\mu\nu}= \frac12 \epsilon^{\mu\nu}{}_{\mu'\nu'} 
\phi^{\mu'\nu'}
\en
and the first order differential equation
\be
\partial_\mu \phi^{\mu\nu}\ = 0\,.
\ee
Using the self-duality condition to eliminate $\phi^{ij}$ in favor of 
$\phi^{kd}$,
the first order equation becomes
\be
\partial_d \phi^{id}
=  -\epsilon^{ij}{}_{k} \partial_j   \phi^{kd}\,.
\ee
If we know $\phi$ on a surface $x^{d}=\mathit{const}$, then we can 
integrate this first order equation to get $\phi$ everywhere.
Self-dual 2-forms in 4d are thus Cauchy fields.
Our goal is to identify the other fields of this type
in all dimensions $d>2$.

\subsection{Level 1 short representations}

Differential equations for $\phi$ correspond
to shortened representations of the conformal algebra $\so(d,2)$.
In the radial quantization of a CFT, each conformal primary field
$\phi(x)$ corresponds to an irreducible lowest weight representation
of the conformal algebra $\so(d,2)$.
The lowest weight $\Delta$ is the lowest eigenvalue of the dilation 
generator $D$.
The
lowest weight states $|\phi\rangle$ form an irreducible representation $V$ of the 
euclidean rotation Lie algebra $\so(d)$.
The field $\phi(x)$ is a vector-valued field of spin $V$ and conformal weight (or
scaling dimension) $\Delta$.
The conformal transformation properties of $\phi(x)$ are completely 
determined by the data $(V,\Delta)$.

The field $\phi(x)$ satisfies a first order
differential equation iff the states $|\phi\rangle$
are annihilated by some linear combination of the
translation generators $P_\mu$. The $P_\mu$ act as
raising operators in the conformal algebra,
adding 1 to the weight.
If a linear combination of the $P_{\mu}$
annihilates $|\phi\rangle$, this means that there
is a null subspace of states at level 1 in the representation.
The representation is said to be a {\it level 1 short
representation}.
In $d=2$ for instance, the Cauchy-Riemann differential equation
(\ref{Cauchy2d}) comes from the null state condition ${\bar
L}_{-1}|\phi\rangle=0$.  For general $d$, classifying the primary fields that satisfy
first order differential equations is equivalent to classifying the
level 1 short representations of the conformal algebra.

The inner product matrix of the conformal representation at level 1 is
easily computed in terms of the commutation relations of the conformal
algebra.  There is a null state on level 1 iff the determinant of the
level 1 inner product matrix equals 0.  This is a polynomial equation
on the conformal weight $\Delta$ with coefficients that depend on the
$\so(d)$ representation $V$.
This type of argument is of course very familiar
from the derivation of unitarity bounds. A necessary
condition for the conformal representation to be
unitary is that the inner product matrix at level 1 be
non negative. For a given $\so(d)$ representation $V$,
the inner product matrix is positive definite for large values of $\Delta$.
As $\Delta$ decreases, it eventually reaches a value where
one or more null states appear.
That is the unitarity bound on $\Delta$.  Below that value of 
$\Delta$, the level 1 inner product matrix has at least one negative 
eigenvalue.
Thus, if we insist on unitarity, the only
level 1 short representations that can occur are at the level 1 
unitarity bound.
If we do not require unitarity, more possibilities
open up, as there are in general several values of
$\Delta$ where null states appear on level 1.
The conserved spin 1 current discussed above does have null
states at level 1, which leads to its conservation equation,
but it does not have the Cauchy property.
The central question of this work is thus: 
which level 1 null states give Cauchy differential equations?

\subsection{Summary of results}
Let us summarize our results.
For unitary fields, we find a complete answer to this question.

In even dimensions $d=2n$,
a unitary Cauchy field
is a conformal field
whose spin is an $\so(d)$ representation
$V_{s}$ with highest weight $\lambda_{s}$,
\eq\label{Vseven}
V_{s} = V_{\lambda_{s}}
\,,\qquad
\lambda_{s} = (|s|,\ldots,|s|,s)
\,,\quad
s \in \frac{1}{2}\mathbb{Z} 
\,\ ,
\en
and whose conformal weight is
\eq
\Delta_s =
\left\{
\begin{array}{ll}
n-1+|s| \quad & s\neq0 \\
0 & s= 0 
\,.
\end{array}
\right .
\en
The case $s=0$ is the trivial case $\phi=1$,  the identity 
operator, which of course satisfies the 
first order equation $\partial_{\mu}\phi=0$.

In odd dimensions $d=2n+1$, the unitary Cauchy fields
are the primary fields  $(V_s,\Delta_s)$
\be\label{Vsodd}
V_{s} = V_{\lambda_{s}}
\,,\qquad
\lambda_{s} = (s,\ldots ,s)
\,,\quad
\text{with }\ s = 0\ \text{ or }\ s=\frac{1}{2}
\,,
\ee
\eq
\Delta_0 = 0\,,\qquad \Delta_{1/2} = n-\frac12
\,.
\en
The case $s=0$ is again the identity operator.

We prove that these lists exhaust all unitary Cauchy fields
in $d>2$.

It was shown in  \cite{Siegel:1988gd}
that 
the conformal
primary fields which
satisfy the massless Klein-Gordon equation
\be\label{KG}
\partial^{\mu}\partial_{\mu} \phi(x) = 0\ 
\ee
consist exactly of  the free massless scalar field
and the fields $(V_s,\Delta_s)$.
For this reason the fields in representations $V_s$ were called ``free''
in \cite{Siegel:1988gd}. This was a misnomer,
since the Klein-Gordon equation did not immediately imply
that the correlation functions factorize as free-field correlation 
functions into products of 2-point functions
according to the Wick contraction rule.
For example, non-abelian current algebras in $d=2$, and, more 
generally, $W$-algebras,
satisfy the Klein-Gordon equation (\ref{KG}) but 
are certainly not free.\footnote{For scalar fields with
$\partial^{\mu}\partial_{\mu} \phi(x) = 0$ in $d>2$ 
factorization was argued in \cite{Dymarsky:2015jia}.
The argument assumes however that the conformal Coleman-Mandula
theorem (see below) holds. We thank Sasha Zhiboedov for discussion of this
and related matters.}

Here we prove that all unitary Cauchy fields in $d>2$ are indeed free fields
by using the Cauchy property to constrain 
the possible modes of the field. 
From the mode expansions, we find constraints on the singular part of 
the operator product expansion of the field with its adjoint field
which imply, for $d>2$, that the commutators of the modes are multiples of the 
identity operator,
which  establishes
that indeed the unitary Cauchy fields all have free-field
correlation functions. We find that only the massless spinor and the self-dual 
$n$-form field can have a local energy-momentum tensor, in accord with
the Weinberg-Witten theorem \cite{Weinberg:1980kq}.

It would have been very nice to find  non-free Cauchy fields that could be 
used to construct non-trivial conformal field theories.
In a sense however the negative result is not unexpected, as it is related
to results on conserved higher spin currents in quantum
field theory in $d>2$. 
Note that short representations lead to conserved currents: 
suppose we have
a field $\phi_a(x)$ in a representation $V$ of $\so(d)$, with
$a$ being an index for $V$,
and suppose that it
satisfies a first-order differential equation
$A^{\mu a}_b\partial_\mu\phi_a(x)=0$.
Then $J^\mu_b=A^{\mu a}_b\phi_a(x)$ is
a conserved current. 
$J$ will be a higher spin conserved current
for all but the smallest representations $V$.
Note that this is somewhat different from
the usual construction of higher spin currents as
bilinears, since here $J$ is linear in the
underlying fields.
Nonetheless, the Coleman-Mandula theorem \cite{Coleman:1967ad}
states that higher spin conserved currents in $d>2$ dimensions must be free.
The original theorem only applies to theories with a mass gap,
so it is not directly applicable here. 
It is believed however that a similar theorem also holds for 
conformal field theories. There has been recent progress
towards proving such a conformal Coleman-Mandula theorem,
although it seems that nothing to date has been
fully proven.
It was argued in \cite{Maldacena:2011jn} for conserved
higher spin currents in $d=3$, in \cite{Alba:2013yda} 
for symmetric traceless fields in $d=4$, and in \cite{Stanev:2013qra}
for 4-, 5- and 6-pt functions of the
energy-momentum tensor in $d=4$.
For higher dimensions there has been work 
\cite{Boulanger:2013zza}
on the construction of higher spin algebras,
arguing that for $d=3$ and $d>5$ under certain assumptions
free theories are the only
such higher spin algebras.
From this point of view, we are proving here a partial Coleman-Mandula theorem in arbitrary dimension.

In the final section, we show that the contour deformation technique 
can be used for any Cauchy conformal field, in any dimension $d$.  We 
show that there always exist enough smearing 
functions to capture all the modes of the field on a surface of 
codimension 1.

Our results for non-unitary conformal Cauchy fields are incomplete.
We classify all possible spins and scaling dimensions for $d=3$ 
and 4.
For $d>4$, we find a restricted list of possible spins and scaling 
dimensions, but we do not prove that all the possibilities on the 
list do in fact have the Cauchy property.
We derive the mode expansions only for the non-unitary cases 
where the spin is one of the $\so(d)$ representations $V_{s}$ but 
the scaling dimension is not the unitary value.
We do not show that the non-unitary theories are free.
It is thus possible that
there are interesting interacting non-unitary Cauchy fields in 
dimensions $d>2$, whose properties could be investigated
by a straightforward application of the methods used here.

\section{The conformal algebra and conformal fields}

\subsection{The conformal algebra and its representations}
We consider euclidean CFTs in the radial quantization.
The conformal operator algebra is $\so(d,2)$. 
It has $(d+2)(d+1)/2$ generators:
the generators $P_{\mu}$ of translations,
the generators $L_{\mu\nu}$ of $\so(d)$, the euclidean rotations,
the generator $D$ of dilations,
and the generators $K_{\mu}$  of the special conformal transformations.
The commutation relations are
\begin{gather}
[P_{\mu},\,P_{\nu}]=0
\,,\qquad
[K_{\mu},\,K_{\nu}]=0
\,,\qquad
[K_{\mu},\,P_{\nu}]= 2 \delta_{\mu\nu} D - 2 L_{\mu\nu}
\nonumber\\
[D,\,P_{\mu}] = P_{\mu}
\,,\qquad
[D,\,L_{\mu\nu}] = 0
\,,\qquad
[D,\,K_{\mu}] = -K_{\mu}
\label{eq:fundamental}
\\
[L_{\mu\nu},\,P_{\sigma}] = \delta_{\nu\sigma}P_{\mu} - \delta_{\mu\sigma}P_{\nu}
\,,\qquad
[L_{\mu\nu},\,K_{\sigma}] = \delta_{\nu\sigma}K_{\mu} - \delta_{\mu\sigma}K_{\nu}
\nonumber\\
[L_{\mu\nu},\,L_{\rho\sigma}] = \delta_{\nu\rho}L_{\mu\sigma}
- \delta_{\nu\sigma}L_{\mu\rho}
+ \delta_{\mu\sigma}L_{\nu\rho}
- \delta_{\mu\rho}L_{\nu\sigma}
\nonumber
\end{gather}
The adjointness relations are
\eq
P_{\mu}^{\dagger} = K_{\mu}
\,,\qquad
D^{\dagger} = D
\,,\qquad
L_{\mu\nu}^{\dagger} = -L_{\mu\nu}
\,.
\label{eq:adjointness}
\en
Our generators differ by a factor of $i$ from the usual ones in the 
physics literature.

The conformal generators implement the conformal vector fields:
\eq
\begin{array}{cc}
\text{generator} & \text{vector field}\\[1ex]
P_{\mu} \qquad & \partial_{\mu} \\
D       \qquad &  x^{\mu}\partial_{\mu} \\
L_{\mu\nu} \qquad & (\delta^{\sigma}_{\mu} x_{\nu} - 
\delta^{\sigma}_{\nu}x_{\mu})\partial_{\sigma}\\
K_{\mu} \qquad & (2 x_\mu x^\sigma- x^2 \delta_{\mu}^{\sigma}) \partial_\sigma
\,.
\end{array}
\en
The ground state is annihilated by all the conformal generators, so 
the correlation functions are invariant under the complex conformal 
algebra $\so(d+2,\Complexes)$.  The conformal symmetries of euclidean 
space form the real subalgebra $\so(d+1,1)$.

The space of states of the radial quantization is constructed from 
the correlation functions with respect to the reflection in the unit 
sphere, the inversion $R: x^{\mu}\mapsto x^{-2} x^{\mu}$.  
Writing $X[v]$ for the operator generator implementing the conformal vector 
field $v$,  the adjoint is $X[v]^{\dagger} = -X[Rv]$,
thus the adjointness relations of (\ref{eq:adjointness}) 
above.
Extending $X[v]$ to complex vector fields $v$,
the vector fields satisfying $Rv = \bar v$
have skew-adjoint generators.
The usual Minkowski space quantization is constructed with respect to 
the reflection in the hyperplane $x^{d}=0$,
$R_{\mathit{Mink}}: (\vec x, x^{d}) \mapsto (\vec x, -x^{d})$.
So, in the Minkowski space quantization, $P_{d}^{\dagger}=P_{d}$ and 
$P_{i}^{\dagger}=-P_{i}$ (recall that our generators differ from the usual 
ones by a factor of $i$).
The two reflections, $R$ and $R_{\mathit{Mink}}$ are conjugate to 
each other in the euclidean conformal group, so the
two Lie algebras of skew-adjoint generators 
are not the same, but they are isomorphic to each other, both
isomorphic to $\so(d,2)$.

A conformal field theory  has a complete set of scaling fields $\phi_{i}(x)$ with scaling dimensions 
$\Delta_{i}$, satisfying
\eq\label{confrad}
[P_{\mu},\,\phi_{i}(x)] = \partial_{\mu}\phi_{i}(x)
\,,\qquad
[D,\,\phi_{i}(x)] = (x^{\mu}\partial_{\mu}+ \Delta_{i}) \phi_{i}(x)
\,.
\en
The radial quantization gives a space of states in one-to-one 
correspondence with the scaling fields.
The operator-state correspondence maps the scaling field $\phi_{i}(x)$ to
an eigenstate  of the dilation generator $D$,
\eq
\phi_{i}(x) \leftrightarrow \ket{\phi_{i}} = \phi_{i}(0) \ket{0}
\,,\qquad
D \ket{\phi_{i}} = \Delta_{i} \ket{\phi_{i}}\ ,
\en
where $\ket{0}$ is the ground state, corresponding to the identity 
field 1.
The ground state $\ket{0}$
is annihilated by all the conformal generators.
Correlation functions are given by ground state expectation values
of radially ordered products of fields.

The generators of the conformal algebra act as operators on
the state space.
The eigenvalue of $D$ is the conformal weight.
The generators $P_{\mu}$ raise the weight by 1, and
the generators $K_{\mu}$ lower the weight by 1.
The conformal lowest weight states are the states that are
killed by the lowering operators $K_{\mu}$,
\eq
K_{\mu} \ket{\phi} = 0
\,,
\qquad
D \ket{\phi} = \Delta \ket{\phi}
\,.
\en
The full space of states is generated from the lowest weight states
by the action of the raising operators $P_{\mu}$.
The $\so(d)$ generators $L_{\mu\nu}$ commute with $D$, so they take
lowest weight states to lowest weight states.  The space of lowest
weight states thus decomposes into a sum of irreducible
representations of $\so(d)$.  We write $\ket{\phi}$ for the finite 
dimensional vector space of lowest
weight states of weight $\Delta$ in an irreducible $\so(d)$ representation $V$.  The
raising operators $P_{\mu}$ acting on $\ket{\phi}$ generate an
irreducible lowest weight representation of the conformal algebra,
with lowest weight $\Delta$.

The conformal primary fields $\phi(x)$ are in one-to-one 
correspondence with the conformal
lowest weight states $\ket{\phi}$ and are labeled by the same data
$(V,\Delta)$.
The representation $V$ is called the {\it spin} of
the field $\phi(x)$.
Writing $\phi_{a}(x)$ for the component fields of the representation 
$V$,
the generators $L_{\mu\nu}$ act on the lowest weight states by  
matrices $M_{\mu\nu}$ on $V$,
\eq
L_{\mu\nu} \ket{\phi_{b}}
= M_{\mu\nu}{}^{a}_{b} \ket{\phi_{a}}\ .
\label{eq:LM}
\en
The matrices $M_{\mu\nu}$ satisfy the same commutation relations as the $L_{\mu\nu}$,
\eq
[M_{\mu\nu},\,M_{\rho\sigma}] = \delta_{\nu\rho}M_{\mu\sigma}
- \delta_{\nu\sigma}M_{\mu\rho}
+ \delta_{\mu\sigma}M_{\nu\rho}
- \delta_{\mu\rho}M_{\nu\sigma}
\,.
\en
There is a unique hermitian inner product on the representation $V$ 
such that
\eq
M_{\mu\nu}^{\dagger} = - M_{\mu\nu}
\,.
\en
The inner product on the entire conformal representation is 
the unique inner product determined by the adjointness relations
 (\ref{eq:adjointness}).

The action of the conformal generators on the  conformal fields
can be derived from the operator state correspondence, the action of 
the conformal generators on the lowest weight states, and 
the translation covariance
\eq
\phi(x) = e^{x^{\mu}P_{\mu}} \phi(0) e^{-x^{\mu}P_{\mu}} 
\,,\qquad
\phi(x) \ket0 =  
e^{x^{\mu}P_{\mu}}\ket\phi
\,.
\en
The results are
\aeq{
[P_{\mu},\,\phi(x)] &= \partial_{\mu}\phi(x)\,, \\
[D,\,\phi(x)] & = x^{\mu}\partial_{\mu}\phi(x) + \Delta \phi(x)\,,
\\
[L_{\mu\nu},\,\phi_{b}(x)] & = (x_{\nu}\partial_{\mu} - 
x_{\mu}\partial_{\nu}) \phi_{b}(x) + M_{\mu\nu}{}^{a}_{b} \phi_{a}(x)\,,
\label{eq:Lmunuphi}
\\
[K_{\mu},\,\phi_{b}(x)] & = (2 x_\mu x^\sigma- x^2 \delta_{\mu}^{\sigma}) \partial_\sigma  \phi_{b}(x)
+ 2 x^{\sigma} ( \Delta \delta_{\mu\sigma} \delta^{a}_{b}-M_{\mu\sigma}{}^{a}_{b}) 
\phi_{a}(x)\,.
}

\subsection{Level 1 short representations}\label{ss:level1}
Let us now study the irreducible lowest weight representations of the conformal 
algebra in more detail.
Let $\phi(x)$ be a conformal field of dimension $\Delta$ in the $\so(d)$ representation 
$V$. As we argued above, the full representation is obtained
by acting with raising operators $P_\mu$. For a generic representation,
we expect all those {\it descendent} states to be linearly independent.
Such a representation is called {\it long}. We are interested
in {\it short}  or {\it degenerate} representations, where
there are linear relations among some of the descendent states.

In mathematical language, the {\it Verma module} is the vector
space spanned by all the formal products
$P_{\mu_{1}}\cdots P_{\mu_{N}}\ket{\phi_{a}}$ symmetric in the indices
$\mu_{i}$.
We have a linear map from the Verma module to the physical 
state space, taking the formal product to the physical product of 
operators acting on the lowesst weight state.
The kernel of this linear map is the space of null states --- the 
linear relations in the Verma module.

The weight space $D = \Delta +N$ is 
called {\it level $N$}.
Let us consider level 1,
spanned by the states $P_{\mu}\ket{\phi_{a}}$.
Suppose the conformal representation is degenerate on level 1.
Then there are identities of the form
\be
A^{\mu b}_{a} P_{\mu}\ket{\phi_{b}}=0\,.
\label{eq:1storderdiffeqn1}
\ee
This is equivalent to
\eq
A^{\mu b}_{a} \partial_{\mu}\phi_{b}(0)\ket{0}=0
\label{eq:1storderdiffeqn2}
\en
which, by a standard quantum field theory argument from translation 
invariance of correlation functions, is equivalent to
\eq
A^{\mu b}_{a} \partial_{\mu}\phi_{b}(x) = 0 \,,
\label{eq:1storderdiffeqn3}
\en
a first order differential equation with constant coefficients.

Conversely, if $\phi(x)$ satisfies 
a first order differential equation with constant coefficients,
the differential equation can alway be written as a set of 
equations in the form
(\ref{eq:1storderdiffeqn3}),
which implies (\ref{eq:1storderdiffeqn2}) and 
(\ref{eq:1storderdiffeqn1}).
So the conformal representation is degenerate on level 1.

In fact, we do not need the condition of constant coefficients.
If $\phi(x)$ is a conformal field, then a first order differential 
equation with non-constant coefficients is equivalent to a first order 
equation with constant coeffients.
Suppose $\phi$ satisfies 
\eq
A^{\mu b}_{a}(x) \partial_{\mu}\phi_{b}(x) = 0
\label{eq:diffeqnnonconst}
\,.
\en
For each $x$, conjugate with translation operators to get
\eq
A^{\mu b}_{a}(x) \partial_{\mu}\phi_{b}(0) \ket0 = 0
\label{eq:xnull}
\,.
\en
This is equivalent to performing the radial quantization with $x$ as 
origin.
Equation (\ref{eq:xnull}) asserts that a certain subspace $\mathcal{N}(x)$ of 
level 1 states is null.
Let $\mathcal{N}$ be the span of all the $\mathcal{N}(x)$ for all $x$,
\eq
\mathcal{N} = \mathop\oplus_{x} \mathcal{N}(x) \subset \Complexes \otimes V
\,.
\en
All the constraints on $\phi$ from the original differential equation 
(\ref{eq:diffeqnnonconst}) are expressed in the fact that $\mathcal{N}$ is a 
null subspace.  The fact that $\mathcal{N}$ is a null subspace is expressed as 
well by the differential equation with constant coefficients
\eq
P_{\mathcal{N}}{}^{\mu}_{\nu}\partial_{\mu}\phi(x) =0
\,,
\en
where $P_{\mathcal{N}}$ is the projection on the subspace 
$\mathcal{N}$.
So the original first order differential equation on $\phi(x)$ with non-constant 
coefficients is equivalent to a differential equation with constant 
coefficients.

Therefore we can say that the conformal field $\phi(x)$ satisfies a
first order differential equation iff the conformal representation is
degenerate on level 1.

To see when level 1 is degenerate, we make the standard calculation 
of the matrix of inner products of the level 1 states,
\aeq{
\bra{\phi_{b'}}P_{\nu'}^{\dagger} P_{\nu} \ket{\phi_{b}}
&= \bra{\phi_{b'}}K_{\nu'} P_{\nu} \ket{\phi_{b}}
= \bra{\phi_{b'}}[K_{\nu'}, \,P_{\nu}] \ket{\phi_{b}}\\
&
= \bra{\phi_{b'}} (2 \delta_{\nu'\nu} D - 2 L_{\nu'\nu})  \ket{\phi_{b}}
\\
&= \bra{\phi_{b'}} (2 \delta_{\nu'\nu} \Delta \delta_{b}^{a} - 2 
M_{\nu'\nu}{}^{a}_{b})  \ket{\phi_{a}}
\label{eq:inprod}
}
The formal states $P_{\mu}\ket{\phi_{a}}$ of the Verma module form 
the $\so(d)$ 
representation $\Complexes^{d} \otimes V$, where $\Complexes^{d}$ is 
the fundamental representation.
Define the self-adjoint matrix $\hat M$ on $\Complexes^{d} \otimes V$,
\eq
\hat M^{\mu a}_{\nu b} =   M^{\mu}{}_{\nu}{}^{a}_{b}
\,.
\en
Then the inner product on level 1 states given by equation (\ref{eq:inprod}) is the hermitian quadratic form on
$\Complexes^{d} \otimes V$ corresponding to the self-adjoint matrix
$2(\Delta - \hat M)$.
Therefore the representation is degenerate on level 1 iff $\Delta$ is 
an eigenvalue of $\hat M$ \cite{Minwalla:1997ka}.  The null vectors 
are then the eigenspace 
$\hat M = \Delta$.

A primary field $\phi(x)$ thus satisfies a first order differential 
equation  iff $\Delta$ is 
an eigenvalue of the self-adjoint matrix $\hat M$ acting on 
$\Complexes^{d} \otimes V$.

The conformal representation is unitary on level 1 iff $\Delta - \hat 
M\ge 0$ which is to say that $\Delta$ cannot be smaller than the largest eigenvalue 
of $\hat M$.  The largest eigenvalue of $\hat M$ is the level 1 {\it 
unitarity bound}, the lower bound on $\Delta$ imposed by
unitarity.

For example, consider 
a scalar field --- a field in the trivial representation of $\so(d)$.
The trivial representation has $\hat M = 0$, so the level 1 
unitarity bound is at $\Delta = 0$.
At the unitarity bound, at $\Delta =0$, the level 1 
matrix of inner products is identically zero, so all the states on 
level 1 are null, so the field satisfies the first order differential 
equation $\partial_{\mu}\phi(x) = 0$,
which implies that $\phi=1$, the identity field.

For scalar fields,
there is an additional unitarity condition at level 2,
\be
\Delta(\Delta-(d-2)/2) \geq 0\ .
\ee
This of course allows the $\Delta=0$ identity field -- the ground state -- but 
it forces any non-trivial scalar field to have dimension at least that of the free 
scalar field, $(d-2)/2$.

The complete  set of unitarity conditions, taking account of all levels of 
the conformal representation, are known.
For scalar fields, the level 1 and level 2 conditions are necessary 
and sufficient for unitarity.
For fields in a non-trivial representation $V$,
the level 1 unitarity condition is necessary and sufficient.
This was established first for $d=3$ \cite{Evans:1967} and $d=4$ \cite{Mack:1975je}
and, finally, for any dimension \cite{MR733809}.

\subsection{The matrix $\hat M$ in terms of Casimir invariants}
Since finding the eigenvalues of the matrix $\hat M$ is the
central issue, let us rewrite it in terms of basic
Lie algebra theoretic objects.
Normalize the quadratic Casimir invariant of a representation 
$V$ of $\mathbf{so}(d)$ as
\eq
\label{eq:Casimir}
C_{2}^{d}(V) = - \frac14 M_{\mu\nu}M^{\mu\nu}
\,.
\en
The matrix $\hat M$ on $\Complexes^{d} \otimes V$ is then simply
\eq
\label{eq:MhatCasimirs}
\hat M = \mathbf1\otimes C_{2}^{d}(V) + C_{2}^{d}(\Complexes^{d})  \otimes \mathbf1
- C_{2}^{d}(\Complexes^{d} \otimes V)
\,.
\en
To see this, note that the  fundamental representation, which is 
given in 
equation (\ref{eq:fundamental}), is generated 
by the matrices
\eq
M^{F}_{\mu\nu}{}^{\rho}_{\sigma} = 
\delta_{\mu}^{\rho}\delta_{\nu\sigma} - \delta^{\rho}_{\nu}\delta_{\mu\sigma}
\en
so
\eq
M^{F}_{\mu\nu}{}^{\rho}_{\sigma}\, M^{\mu\nu}{}^{a}_{b}
= 2 M^{\rho}{}_{\sigma}{}^{a}_{b}
=  2 \hat M^{\rho a}_{\sigma b}
\en
so
\eq
\hat M = \frac12 M^{F}_{\mu\nu} \otimes M^{\mu\nu}
= \mathbf1\otimes C_{2}^{d}(V) + C_{2}^{d}(\Complexes^{d})  \otimes \mathbf1
- C_{2}^{d}(\Complexes^{d} \otimes V)
\,.
\en
So the eigenvalues of $\hat M$ are gotten by decomposing 
$\Complexes^{d}\otimes V$ into irreducible components and finding 
their quadratic Casimir invariants.
The classical representation theory needed for this is collected in 
the appendices.  
We postpone calculating until we have found how to tell which $\mathbf{so}(d)$ 
representations $V$ give
degenerate conformal representations that give first order differential 
equations with the Cauchy property.

\section{First order differential equations and the Cauchy property}
\subsection{The first order differential equation}

A conformal field $\phi(x)$ of spin $V$ satisfies a first order differential 
equation iff $\Delta$ is an eigenvalue of $\hat M$.
We want to know for what $\mathbf{so}(d)$ representations $V$ and scaling dimensions $\Delta$ 
does this first order differential equation have the Cauchy property.

The central object for that analysis is $\hat P_\Delta$, the projection matrix acting on $\Complexes^{d}\otimes V$ 
that projects on the eigenspace $(\Complexes^{d}\otimes V)_{\Delta}$ 
where $\hat M = \Delta$,
\be
\hat P_\Delta : \Complexes^{d}\otimes V \rightarrow (\Complexes^{d}\otimes V)_{\Delta}\ .
\ee
To simplify notations, we will often just write $\hat P$ for $\hat P_\Delta$.
The matrix elements of $\hat P$ are $\hat P ^{\mu r}_{\nu s}$.
The eigenspace $(\Complexes^{d}\otimes V)_{\Delta}$ is the space of null states on level 1, so
\eq
\hat P ^{\mu a}_{\nu b} P_{\mu} \ket{\phi_{a}} = 0
\,.
\en
Suppressing the indices for $V$, write
$\hat P ^{\mu}_{\nu}$ for the matrix on $V$
with matrix elements $\hat P ^{\mu a}_{\nu b}$.
Then the null state conditions are written
\eq
\hat P ^{\mu}_{\nu} P_{\mu} \ket{\phi} = 0
\,,
\en
equivalent to the differential equation
\eq
\label{eq:Cauchydiffeqn}
\hat P ^{\mu}_{\nu} \partial_{\mu} \phi(x) = 0
\,.
\en

\subsection{The Cauchy property as an algebraic condition}\label{ss:CauchyAlgebraic}
The Cauchy property is the condition on the differential equation that the values of $\phi(x)$ on a 
codimension 1 submanifold completely determine $\phi(x)$ everywhere.
For a rotationally invariant first order equation with constant 
coefficients,
this is simply the condition that $\partial_d\phi(x)$ is completely determined
by the $\partial_{i}\phi(x)$, which is an algebraic condition.

Suppose $\phi(x)$ and $\tilde\phi(x)$ have the same spatial 
derivatives at $x$, $\partial_{i}\phi(x) =\partial_{i} \tilde \phi(x)$.
We need the differential equation to imply that 
$\partial_{d}\phi(x) =\partial_{d} \tilde \phi(x)$.
Writing $\delta \phi = \tilde \phi -\phi$,
we need
\eq
\partial_{i}\delta\phi(x) = 0 
\implies
\partial_{d}\delta\phi(x) = 0 
\,.
\en
Both $\phi(x)$ and $\tilde \phi(x)$ satisfy the differential 
equation (\ref{eq:Cauchydiffeqn}) and the equation is linear, so $\delta\phi(x)$ also 
satisfies it,
\eq
0 = \hat P ^{\mu}_{\nu} \partial_{\mu} \delta \phi(x)
\,.
\en
Writing the indices for $V$ explicitly, this is
\eq
\hat P ^{da}_{\nu b} \partial_{d} \delta \phi_{a}(x)
+\hat P ^{ia}_{\nu b} \partial_{i} \delta \phi_{a}(x)
= 0
\,.
\en
But the spatial derivatives are zero, so the differential equation 
gives
\eq
 \hat P ^{da}_{\nu b} \partial_{d} \delta \phi_{a}(x) =0
\,.
\en
We need this to imply that $\partial_{d}\delta \phi_{r}(x)=0$.

Define
$\hat P^{d}$ to be the matrix with matrix elements
$\hat P ^{da}_{\nu b}$.
It is a linear map from $V$ to the $\hat M=\Delta$ eigenspace,
\eq
\hat P^{d} : V \rightarrow (\Complexes^{d}\otimes V)_{\Delta}
\,,\qquad
\hat P^{d} v = \hat P(\hat e_{d}\otimes v)
\,,
\en
where $\hat e_{d}$ is the unit vector in the $d$-direction in 
$\Complexes^{d}$.

We need
\eq
\hat P^{d} \partial_{d} \delta \phi(x) = 0
\implies
\partial_{d} \delta \phi(x) = 0
\,. 
\en
But $\partial_{d} \delta \phi(x)$ can be any vector in $V$.
So this is simply the condition
\eq
\forall v\in V\,,\; \hat P^{d} v = 0 \implies v=0
\,,
\en
which is the condition that $\hat P^{d}$ is injective.
So we have shown that the Cauchy property is equivalent to the 
condition that $\hat P^{d}$ is injective,
\begin{itemize}
\item[\Aone] The first order differential equation satisfied by $\phi(x)$ has
the Cauchy property iff the matrix $\hat P^{d}:V\rightarrow (\Complexes^{d}\otimes V)_{\Delta}$ 
is injective.
\end{itemize}
\noindent
Now define $\hat P^{d}_{d}$ to be the matrix with matrix elements 
$\hat P ^{da}_{d b}$. It is a linear map from $V$ to $V$,
\eq
\hat P^{d}_{d}: V \rightarrow V 
\,,\qquad
\hat P^{d}_{d}v = \mathrm{Proj}_{\Complexes\hat e_{d}\otimes V}
(\hat P^{d} v)
\,.
\en
We will argue that \Aone is equivalent to
\begin{itemize}
\item[\Atwo] The first order differential equation satisfied by $\phi(x)$  has
the Cauchy property iff the matrix $\hat P^{d}_{d}: V \rightarrow V $ 
is invertible.
\end{itemize}
To see the equivalence,
first note that if $\hat P^d$ is not injective, then there is a 
non-zero
vector $v$ with $\hat P^d v = 0$, which implies $\hat P^d_d v = 0$ ,
so $\hat P^{d}_{d}$ cannot be  invertible.
Now suppose $\hat P^{d}_{d}$ is not invertible.  We will use the 
invariant inner-products on $V$ and on the tensor product space
$\Complexes^{d}\otimes V$.
If $\hat P^{d}_{d}$ is not invertible, there must be
a non-zero $v\in V$ such that $\hat P^{d}_{d} v = 0$, so
\eq
(v,\hat P^{d}_{d} v)=0
\,.
\en
But $(v,\hat P^{d}_{d} v)=(\hat e_{d}\otimes v, \hat P \hat 
e_{d}\otimes v)$, so
\eq
(\hat e_{d}\otimes v, \hat P \hat e_{d}\otimes v)=0
\,.
\en
$\hat P$ is a self-adjoint projection, so
\eq
(\hat P  \hat e_{d}\otimes v, \hat P \hat e_{d}\otimes v)
= (  \hat e_{d}\otimes v, \hat P^{2} \hat e_{d}\otimes v)
= (\hat e_{d}\otimes v, \hat P \hat e_{d}\otimes v) = 0
\,.
\en
The invariant inner-product is positive definite, so we have
\eq
\hat P^{d} v =  \hat P \hat e_{d}\otimes v = 0
\,,
\en
so we have shown that $\hat P^{d}_{d}$ not invertible implies $\hat 
P^{d}$ not injective.  So the algebraic Cauchy conditions are 
equivalent.

The {\it symbol} of the first order differential operator 
$\hat P ^{\mu}_{\nu} \partial_{\mu}$
in the differential equation (\ref{eq:Cauchydiffeqn}) is the map from 
unit co-vectors to matrices,
\eq
\hat n_{\mu} \mapsto \hat n_{\mu}\hat P ^{\mu a}_{\nu b} 
\,.
\en
By rotational invariance, we might as well choose $\hat n$ in the 
$d$-direction, in which case the symbol is the matrix $\hat P^{d}$.
So the symbol for any value of $\hat n_\mu$ is conjugate to $\hat 
P^{d}$ by some rotation.
So our Cauchy condition is exactly the condition that the symbol of 
the first order differential operator  is injective.

A differential operator is said to be {\it elliptic} when its symbol
is invertible.  A differential operator is said to be {\it
overdetermined elliptic} when its symbol is injective but not 
invertible.
(Sometimes the label {\it elliptic}  is used for both.)

It seems to us that the natural 
generalization of $\bar \partial$ to conformal fields in $d>2$ is 
the Cauchy property as we have defined it, the property of 
having an injective symbol, which is the property that allows
integrating the first order differential equation 
from data on a codimension 1 submanifold.

The label {\it overdetermined} refers to the fact that not all initial 
data on the codimension 1 submanifold is possible.  
The differential equation imposes linear constraints on the value of 
$\phi(x)$ on the submanifold.
We discuss those constraints in section~\ref{s:overdet} below.

As an illustration, let us explain in more detail how the first order equation 
satisfied by a conserved current, $\partial_{\mu}j^{\mu}(x)=0$, fails 
to satisfy the conditions \Aone and \Atwo.  The first order differential equation 
(\ref{eq:Cauchydiffeqn}) is
\eq
\hat P ^{\mu\alpha}_{\nu\beta} \partial_{\mu} j_{\alpha}(x) = 0
\en
with the projection $\hat P$ being
\eq
\hat P ^{\mu\alpha}_{\nu\beta} = \delta^{\mu\alpha} \delta_{\nu\beta}
\,.
\en
Then
\eq
\hat P^{d\alpha}_{\nu\beta} = \delta^{d\alpha} \delta_{\nu\beta}
\,,\qquad
(\hat P v)_{\nu\beta} = \hat P^{d\alpha}_{\nu\beta} v_{\alpha} = 
v^{d} \delta_{\nu\beta}
\,.
\en
If $v^{d}=0$ then $\hat P^{d}v =0$.  So $\hat P^{d}$ is not injective.
The matrix $\hat P^{d}_{d}$ is
\eq
P^{d\alpha}_{d\beta} = \delta^{d\alpha} \delta_{d\beta}
\en
which is the projection matrix on the $d$-direction, which is 
obviously
not invertible.

\section{Cauchy representations}

From our discussion in the previous section,
we define a field $\phi$ to be a Cauchy field if it satisfies
a first order equation with injective symbol.
This property then determines $\phi$ everywhere uniquely
once we know its value on some codimension 1 submanifold.

We argued in section~\ref{ss:level1} that short representations
lead to differential equations. However, as we illustrated
with the conserved current, not every short representation
leads to a differential equation which satisfies the
Cauchy condition. Cauchy representations are thus a proper
subset of short representations, and we want to classify them. 

In this section, we present all the unitary Cauchy representations.
We show here that all of them have the Cauchy property.  The proof
that there are no other unitary Cauchy representations is given in
appendix~\ref{ss:classifyUnitary}.

\subsection{The identity field}

The most obvious example is the trivial representation.
If $V$ is the trivial $\so(d)$ representation, $M_{\mu\nu}=0$, then
$\Delta=0$ is the level 1 degeneracy condition.
If $\Delta=0$, then all of level 1 is null.
The first order differential equation is
$\partial_{\mu}\phi(x) =0$,
whose solution is the identity field $\phi(x) =1$.
In the following we will thus only consider 
non-scalar (non-trivial) $\so(d)$ representations $V$.

\subsection{$V_s$ for $d=2n$}
We saw at the beginning that the (anti-)self-dual $n$-forms $\Lambda^{n}_{\pm}$
are Cauchy fields in even dimensions $d=2n$.
Let us now describe a larger set of examples in $d=2n$,
namely the representations (\ref{Vseven}),
\eq
V_{s} = V_{\lambda_{s}}
\,,\qquad
\lambda_{s} = (|s|,\ldots,|s|,s)
\,,\quad
s \ne 0
\,.
\en
In particular,
$V_{\pm\frac12}$ are the chiral spinors, $V_{\pm 1}$ are the 
(anti-)self-dual $n$-forms.

For these examples we can make the rather abstract discussion
in (\ref{ss:CauchyAlgebraic}) much more concrete by giving
very explicit expressions for the projectors $(\hat P_\Delta)^{d}_{d}$.
First note from (\ref{sodtensor})  that 
the tensor product $\Complexes^{d}\otimes V_{s}$ decomposes into exactly two 
components,
\eq
\Complexes^{d}\otimes V_{s} =
V_{\lambda_{s}+\epsilon_{1}} \oplus V_{\lambda_{s}\mp \epsilon_{n}}
\,,\qquad \text{for } s = \pm |s|\,,
\en
so that $\hat M$ has two eigenspaces.
Its eigenvalues can be calculated using its expression in terms of 
quadratic Casimirs (\ref{eq:MhatCasimirs}) and formula 
(\ref{eq:quadraticCasimir}) for the quadratic Casimirs,
\eq
\hat M =
\left \{
\begin{array}{ll}
-|s| & \text{on } V_{\lambda_{s}+\epsilon_{1}} 
\\[1ex]
 n-1+|s| &\text{on } V_{\lambda_{s}\mp \epsilon_{n}} 
\end{array}
\right .\ .
\en
Since $\hat M$ has two distinct eigenvalues, and since the sum of the 
eigenvalues is $n-1$, the projection on the 
subspace $\hat M = \Delta$ is simply
\eq
\hat P_\Delta = \frac{\hat M -  (n-1-\Delta)}{\Delta -  (n-1-\Delta)}
\,.
\en
Since $\hat M^{d}_{d} = 0$ we have
\eq
(\hat P_\Delta)^{d}_{d} = \frac{ -  (n-1-\Delta)}{\Delta -  (n-1-\Delta)}
\en
which is a non-zero multiple of the identity matrix.
So $\hat P^{d}_{d} $ is invertible.
Therefore $(V_{s},\Delta)$ is a Cauchy representation for $\Delta$ either of 
the two eigenvalues of $\hat M$.

The largest eigenvalue of $\hat M$ is $ n-1+|s|$,
which is therefore the level 1 unitarity bound.
In fact, since we are considering $s\neq0$ only, 
it is the exact unitarity bound.
So $V=V_{s}$, $\Delta = n-1+|s|$ gives a unitary Cauchy field,
while $V=V_{s}$, $\Delta = -|s|$ gives a non-unitary Cauchy field.
The Cauchy differential equation (\ref{eq:Cauchydiffeqn}) is
\eq
\label{eq:Vseqn}
\left [ \hat M^{\mu}_{\nu} -  (m-\Delta) \delta^{\mu}_{\nu}\right ]\partial_{\mu}\phi(x) = 0
\,,\qquad
m = \frac12 (d-2)
\,.
\en
For the unitary field, $\Delta = |s|+m$,
the differential equation is
\eq
\left ( \hat M^{\mu}_{\nu} +|s| \delta^{\mu}_{\nu}\right )\partial_{\mu}\phi(x) = 0
\,.
\en

\subsection{$V_s$ for $d=2n+1$}
The Cauchy fields in odd dimensions are even more restricted.
From (\ref{sodtensor}) we see that for $s>1/2$ we have
\eq
\Complexes^{d}\otimes V_{s} =
\left\{
\begin{array}{ll}
V_{\lambda_{s}} \oplus
V_{\lambda_{s}+\epsilon_{1}} \oplus V_{\lambda_{s}- \epsilon_{n}}
\,,\qquad & \text{for } s > 1/2\,,
\\[1ex]
V_{\lambda_{s}} \oplus
V_{\lambda_{s}+\epsilon_{1}}
\,,\qquad & \text{for } s = 1/2
\,.
\end{array}
\right .
\en
Only for $s=1/2$ are there again only
two representations in the tensor product,
so only for $s=1/2$ can we repeat the above argument for the Cauchy 
property.
Indeed, in appendix~\ref{ss:classifyUnitary} we show
that $s>1/2$ gives non-Cauchy fields.

For $s=1/2$, the eigenvalues of $\hat M$ are $\frac12(d-1)$ and $-s$.
The unitarity bound is $\Delta= \frac12(d-1)$.
This is the massless spinor field.
The Cauchy differential equation for either the unitary 
or non-unitary Cauchy field is
\eq
\label{eq:Vseqndodd}
\left [ \hat M^{\mu}_{\nu} -  \left (m-\Delta\right ) \delta^{\mu}_{\nu}\right ]\partial_{\mu}\phi(x) = 0
\,,\qquad
m = \frac12 (d-2)
\,.
\en
For the unitary field, $\Delta = m+|s|$, the differential equation is
\eq
\left ( \hat M^{\mu}_{\nu} +|s| \delta^{\mu}_{\nu}\right )\partial_{\mu}\phi(x) = 0
\,.
\en
It can be shown that this equation is equivalent to the massless Dirac equation.

\subsection{Summary of the Cauchy fields}
\label{s:summaryCauchyfields}
We now have the classification of the non-trivial
Cauchy conformal fields in unitary conformal field 
theories.
Write $m=\frac12(d-2)$.

In even dimensions $d$, the non-trivial Cauchy conformal fields are the
conformal primary fields of spin $V$ and conformal weight $\Delta$ in the set
\eq
\left \{(V,\Delta)=(V_{s},|s|+m): s\in \frac12 \Integers,
\; s\ne 0\right\}
\en
In odd dimensions $d=2n$, the only non-trivial Cauchy conformal 
field is the massless spinor field,
$V=V_{s}$, $s=\frac12$, $\Delta=|s|+m$.

For both $d$ even and for $d$ odd, the Cauchy differential equation is
\eq
\label{eq:Vseqndall}
\left [\hat M^{\mu}_{\nu} -  (m-\Delta) \delta^{\mu}_{\nu}\right ]\partial_{\mu}\phi(x) = 0
\,,\qquad
m = \frac12 (d-2)
\,.
\en
For the unitary cases, 
\eq
\Delta = |s|+m
\,,\qquad
\left ( \hat M^{\mu}_{\nu}+|s| \delta^{\mu}_{\nu}\right )\partial_{\mu}\phi(x) = 0
\,.
\en

\subsection{The spin $V_{s}$ Cauchy fields as ``free fields''}

Conformal fields in the representations $V_{s}$ were previously studied in 
somewhat different context \cite{Siegel:1988gd}, as  a subset of the 
so-called ``free field'' 
conformal representations --- the conformal fields $\phi(x)$ 
satisfying the massless Klein-Gordon equation 
\be\label{CauchyKG}
\partial^{\mu}\partial_{\mu}\phi(x)=0\ .
\ee
Indeed, applying $\partial^{\nu}$ to equation (\ref{eq:Vseqndall}) gives 
$\partial^{\mu}\partial_{\mu}\phi(x)=0$.
Conversely, if $P^{\nu}P_{\nu}\ket{\phi}=0$, then
the identity $[K_{\mu},\,P^{\nu}P_{\nu}]\ket{\phi}=0$ gives exactly 
the non-trivial first order differential equation (\ref{eq:Vseqn}),
except for the case of a scalar field with the scaling dimension of the free 
massless scalar field.
A conformal field with non-zero spin  satisfying equation (\ref{eq:Vseqn}) 
necessarily belongs to an $\mathbf{so}(d)$ representation $V$ such 
that $\hat M$ has at most two eigenvalues.  
So the ``free field'' representations are the  representations 
$V_{s}$ plus the representation of the free massless scalar field.

The label ``free field'' for conformal fields satisfying
$\partial^{\mu}\partial_{\mu}\phi(x)=0$ was a misnomer.
Additional work is needed to show that $\phi(x)$ is actually a 
free field, \ie a field whose correlation functions are gaussian, given in terms 
of its two-point function by Wick contractions. For $d=2$, this
is obviously not the case, as shown by the many non-free examples.
In the following two sections we will 
develop the tools to show that for $d>2$, all unitary Cauchy fields
are indeed free.

\section{Mode expansions}\label{s:modes}

In this and the following section we prove two negative results about the Cauchy conformal 
fields in unitary theories in dimensions $d>2$:
First, they are free fields --- their correlation functions are the products 
of the 2-point function following the free-field Wick contraction rules.
Second, they have a local stress-energy tensor only for $|s|\le 1$.

To establish this result, we go through the following steps:
\begin{enumerate}
\item We construct a mode expansion for the field $\phi(x)$, expanding 
in spherical harmonics on the $(d-1)$-sphere.
\item We use the mode expansion and the branching rules for tensor 
products of $\so(d)$ representations to show that the singular part 
of the operator product expansion contains only the identity field.
\item We then point out that the commutator depends only on the singular part 
of the operator product expansion. It follows that the commutator is a multiple of 
the identity, so that the correlation functions are indeed given by free-field 
Wick contractions.
\item We finally point out that, for $|s|>1$, there is no field in the operator product 
expansion with the spin and scaling dimension of the stress-energy 
tensor.
\end{enumerate}

This is perhaps not the most elegant proof, but  
it has the virtue of efficiency. We deviate from the spirit
of $d=2$ holomorphic fields, since we do not 
calculate the commutators by contour deformation.
In fact we have not yet
established that contour deformation is valid.
We return to the question of deformability in 
section~\ref{sect:deformability}.
It would be more elegant to 
use those techniques:  construct the field modes by smearing 
$\phi(x)$ over a codimension1 surface against smearing functions 
that satisfy the dual Cauchy equation,
then calculate the commutator of the field modes by contour 
deformation.
This is a priori a more powerful technique,
because the commutator so derived would depend only on the part of the operator 
product expansion with singularity at least $O(r^{1-d})$, not on the 
entire singular part.

As it happens, our cruder approach is powerful enough to 
establish the two results. We can easily enforce a subset of the constraints 
implied by the Cauchy differential equation to restrict 
the  modes enough to control the singular part of the operator product 
expansion.  We do not need the precise list of field modes --- some 
of the modes on our list might be zero. We will return to 
this question in section~\ref{ss:nonzero} to show that they are all in
fact non-vanishing. For the moment however our list of 
the possible modes is small enough to give our two negative results 
on the unitary Cauchy fields.

\subsection{Expand in spherical harmonics}

We  describe the operator representation of a Cauchy field 
in the radial quantization by expanding the field in spherical 
harmonics.
Suppose $\phi(x)=\phi(r\hat x)$ is a Cauchy conformal field with spin $V_{s}$
and conformal weight $\Delta$.

The rotation Lie algebra $\so(d)$ acts on the field by the generators
\be
\label{eq:Mtot}
M^{\mathit tot}_{\mu\nu} \phi(x) = (M^{\mathit orb}_{\mu\nu} 
 +M^{\mathit spin}_{\mu\nu} ) \phi(x)
\ee
where, as in (\ref{eq:Lmunuphi}), the  $M^{\mathit spin}_{\mu\nu}$  generate the spin 
representation $V_{s}$ and the orbital generators
\eq
\label{eq:orbitalgenerators}
M^{\mathit orb}_{\mu\nu}  = -(x_{\mu}\partial_{\nu} - x_{\nu}\partial_{\mu})
\en
generate the rotations acting on the unit sphere, on functions of $\hat x$.
The functions on the unit sphere decompose
into irreducible representations of the $M^{\mathit orb}_{\mu\nu}$,
the spherical harmonics,
\eq
L_{2}(S^{d-1}) = \mathop\oplus_{l=0}^{\infty} S_{l}
\,,
\en
where
\be
S_l = V_{(l,0,\ldots,0)}
\ee
is the irreducible representation of $\so(d)$ on traceless symmetric $l$-tensors.

Expand the field $\phi$ in spherical harmonics
\eq
\label{eq:modeexpansion1}
\phi(r\hat x) = \sum_{l=0}^{\infty} Y_l(\hat x)  \phi_{l}(r)
\,,\qquad
Y_{l}\in S_{l}
\,,
\en
\eq
Y_l(\hat x)  \phi_{l}(r) = \frac1{l!} \,
\hat x^{\mu_{1}} \cdots \hat x^{\mu_{k}}\,
\phi_{l}(r)_{\mu_{1} \cdots \mu_{k}}
\,.
\en
The mode $\phi_{l}(r)$ is a traceless symmetric $l$-tensor with 
values in $V_{s}$.  Note that we are suppressing the $V_{s}$ index in 
our labeling of the field $\phi(x)$ and its modes $\phi_{l}(x)$.

The field mode $\phi_{l}(r)$ lies in a sub-representation of $S_{l}\otimes 
V_{s}$.
The next steps are to use the Cauchy equation to put constraints on this 
sub-representation.

\subsection{Constrain the modes using 
$\partial^{\mu}\partial_{\mu} \phi=0$}

Again, applying $\partial^{\nu}$ to equation (\ref{eq:Vseqndall}) gives 
$\partial^{\mu}\partial_{\mu}\phi=0$.
So $\phi(x)$ is a harmonic function, with the usual expansion in 
spherical harmonics.

To be explicit, the laplacian in polar coordinates is
\aeq{
\partial_{\mu}\partial^{\mu} &=
r^{-2} 
\left [(r\partial_{r})^{2} + (d-2) r\partial_{r}  -2 C_{2}^{d,{\mathit orb}}\right]
}
where
\eq
C_{2}^{d,{\mathit orb}} = -\frac14 M^{\mathit orb}_{\mu\nu}M_{\mathit orb}^{\mu\nu}
\en
is the quadratic Casimir operator of the orbital representation.
Applying the laplacian to the mode expansion 
(\ref{eq:modeexpansion1}),
using the quadratic Casimir invariants
\eq
C_{2}^{d}(S_{l}) = \frac12 l (l+d-2)
\en
gives
\eq
(r\partial_{r} -l) (r\partial_{r} +d-2 +l)  \phi_{l}(r) = 0
\,.
\en
The general solution is
\eq
\phi_{l}(r) =  r^l \phi^+_{l} +  r^{-l-2m} \phi^-_{l}
\,.
\en
Recall that we defined $m =\frac{d-2}{2}$.

We can read off the scaling dimensions directly, or use 
(\ref{confrad}),
\eq
\dim(\phi^+_{l}) = \Delta+l
\,,\qquad
\dim(\phi^-_{l}) = \Delta-l-2m
\,.
\en
Now we follow the usual convention of labeling the field modes
by minus the scaling dimension,
\be
\phi_k = \left\{ 
\begin{array}{rllrl}
\phi^+_{-k-\Delta}&= \phi^+_{|k+\Delta-m|-m} 
&{:}\qquad&                  &k+\Delta \leq 0\,,\\[1ex]
                   &0&{:}\qquad&       0 < \!\! \!&k+\Delta <  2m\,, \\[1ex]
\phi^-_{k-2m+\Delta}&= \phi^-_{|k+\Delta-m|-m}&{:}\qquad&  2m   \le \!\!\!&k+\Delta\,,
\end{array}
\right.
\label{eq:modes}
\ee

\eq
k+\Delta \in \Integers 
\,,\qquad
\phi_{k}\in S_{|k+\Delta-m|-m}\otimes V_{s}
\,,\qquad \dim(\phi_{k}) = -k
\,,\qquad
[D,\phi_k] = -k\phi_k
\,.
\en
The mode expansion is now
\be
\label{eq:modeexpansion}
\phi(x) = \sum_{\substack{k+\Delta\in \Integers\\ |k+\Delta-m|\geq m}} r^{-k-\Delta} Y_{|k+\Delta-m|-m}(\hat x)
\phi_{k}\ .
\ee
Next note that the correlation functions must be regular as 
$r\rightarrow0$, so $\phi(x)\ket0$ must be regular,
where $\ket 0$ is
the ground state of the radial 
quantization.
Therefore,
\eq
\phi_{k} \ket 0 = 0
\,,\qquad k > -\Delta
\en
and the operator state correspondence is
\eq
\ket{\phi} = \phi(0) \ket0 = \phi_{-\Delta} \ket0
\,.
\en
We will thus identify the $\phi_k$ with $k \geq 2m -\Delta$
with annihilation modes, and the $\phi_k$ with $k\leq -\Delta$
with creation modes.

\subsection{Constrain the modes using the radial Cauchy equation}
\label{s:constrainthemodesradial}

So far, we have only used the fact that any field satisfying 
$\partial^{\mu}\partial_{\mu}\phi=0$ has a discrete expansion in 
harmonic functions to get the mode expansion
(\ref{eq:modeexpansion}).
The mode
$\phi_{k}$
forms a subrepresentation of 
$S_{|k+\Delta-m|-m}\otimes V_{s}$.
The next step is to show that the radial component
of the first order differential equation
(\ref{eq:Vseqndall}),
\eq
\label{eq:diffeqnradial}
x^{\nu}\left [ \hat M^{\mu}_{\nu} -  (m-\Delta) \delta^{\mu}_{\nu}\right ]\partial_{\mu}\phi(x) = 0
\,,
\en
fixes the quadratic Casimir invariant of $\phi_{k}$.
Then we use the known decomposition of the tensor product
$S_{|k+\Delta-m|-m}\otimes V_{s}$
into irreducible representations
to find that the Casimir invariant of $\phi_{k}$ either does not 
occur in the decomposition, or identifies a unique irreducible.
Thus certain of the modes $\phi_{k}$ must vanish identically,
and each of the rest of the $\phi_{k}$ are in
specific irreducible $\so(d)$ representations.

The representation of $\so(d)$ on $\phi(x)$
is the tensor product of the spin and orbital representations,
as given in equation (\ref{eq:Mtot}).
Using equation (\ref{eq:orbitalgenerators}) for the orbital 
generators,
\be\label{spinorbit}
\frac12 M^{\mathit  orb}_{\mu\nu}M_{\mathit  spin}^{\mu\nu}  \phi_{s}(x)
= x^{\nu}\partial^{\mu} M_{\mu\nu}{}^{r}_{s} \phi_{r}(x) 
= x^{\nu} \hat M^{\mu}_{\nu}{}^{r}_{s}  \partial_\mu \phi_{r}(x)
\,.
\ee
The $\mathbf{so}(d)$ quadratic Casimir operators are
\eq
C_{2}^{d,{\mathit orb}} = -\frac14 M^{\mathit orb}_{\mu\nu}M_{\mathit orb}^{\mu\nu}
\,,\qquad
C_{2}^{d,{\mathit spin}} = -\frac14 M^{\mathit  spin}_{\mu\nu}M_{\mathit  spin}^{\mu\nu}
\,,\qquad
C_{2}^{d,{\mathit tot}} = -\frac14 M^{\mathit  tot}_{\mu\nu}M_{\mathit  tot}^{\mu\nu}
\,,
\en
so
\eq
\frac12 M^{\mathit  orb}_{\mu\nu}M_{\mathit  spin}^{\mu\nu}= C_{2}^{d,{\mathit orb}} + C_{2}^{d,{\mathit spin}}
-C_{2}^{d,{\mathit tot}} 
\,,
\en
so equation (\ref{eq:diffeqnradial}), the radial component of the Cauchy differential equation, 
can be written
\eq
\left [C_{2}^{d,{\mathit orb}} + C_{2}^{d,{\mathit spin}}
- C_{2}^{d,{\mathit tot}}
-  (m-\Delta) r\partial_{r}
\right ] \phi(x) = 0
\en
Substituting the mode expansion (\ref{eq:modeexpansion}) and taking 
account of the independence of the terms in the expansion, we get
\eq
\left [C_{2}^{d,{\mathit orb}} + C_{2}^{d,{\mathit spin}}
- C_{2}^{d,{\mathit tot}}
+  (m-\Delta) (k+\Delta)
\right ] \phi_{k} = 0
\,,
\en
which, since $\phi_{k}$ lies in 
$S_{|k+\Delta-m|-m}\otimes V_{s}$,
fixes the quadratic Casimir invariant of $\phi_{k}$ to be
\eq
C_{2}^{d,{\mathit tot}} \phi_{k}
=
\left [C_{2}^{d}(S_{|k+\Delta-m|-m}) +C_{2}^{d}(V_{s})  
+  (m-\Delta) (k+\Delta)
\right ] \phi_{k}
\,.
\en
Using
\eq
C_{2}^{d}(S_{l})\phi_{k} = \frac12 l (l +d-2) \phi_{k}
\,,
\en
with
\eq
l = |k+\Delta-m|-m\,,
\en
we get a simple formula for the quadratic Casimir of $\phi_{k}$,
\eq
\label{eq:Casimirphin}
C_{2}^{d,{\mathit tot}} \phi_{k}
=
\left [
C_{2}^{d}(V_{s})  +
\frac12
(
-\Delta^{2}
+  k^{2}
)
\right ] \phi_{k}
\,.
\en

The second step is to compare these Casimir values
with the Casimirs of the irreducible representations
that occur in the decomposition of the tensor product $S_{|k+\Delta-m|-m}\otimes V_{s}$.
That decomposition is given in \cite{defosseux2010}, Props~9.4 and 9.5.
For  $d$ even,
\be\label{Vsfusion}
S_{l}\otimes V_{(|s|,\ldots,|s|,\pm|s|)}= \mathop\oplus_{s'= 
\max(-|s|,-l+|s|)}^{|s|} V_{(l+s',|s|,\ldots,|s|,\pm s')}\ .
\ee
For $d$ odd, the decomposition of  
the tensor product is, for $s=0$ and $s=1/2$,
\be
\label{eq:decompdodd}
S_{l}\otimes V_{(|s|,\ldots,|s|,|s|)}= \mathop\oplus_{s'= 
\max(-|s|,-l+|s|)}^{|s|} V_{(l+s',|s|,\ldots,|s|,|s'|)}\ .
\ee
For $d$ even, the Casimir invariants of the individual components are
\begin{multline}
C_{2}^{d}(V_{(l+s',|s|,\ldots,|s|,\pm s')})  
=
\\
 C_{2}^{d}(V_{s}) + \frac12\left [ (l+s')(l+s'+d-2) + s'^{2} -|s|(|s|+d-2) - s^{2}
\right ]\ .
\end{multline}
The Casimir of the component increases monotonically in 
$s'$, given the inequalities on $s'$ in the decomposition 
(\ref{Vsfusion}).
Therefore the Casimir of $\phi_{k}$ can match the Casimir of at most one 
component, so $\phi_{k}$ will lie in an irreducible representation, 
or will vanish.

The Casimir of the component labelled by $s'$ agrees with the Casimir of $\phi_{k}$ 
given in equation (\ref{eq:Casimirphin}) iff $s'$ satisfies the 
quadratic equation
\eq
(l+s')(l+s'+d-2) + s'^{2} -|s|(|s|+d-2) - s^{2}
= -\Delta^{2}
+  k^{2}
\,,
\en
which can be re-arranged as
\eq
\left[ s' + \frac12(l+m)\right ]^{2}
-\frac14(k-\Delta+m)^{2}
= (|s|+m-\Delta)(\Delta+|s|)
\,.
\en
The rhs vanishes, because
$\Delta$ has one of the two values: $|s|+m$ in the unitary case, $-|s|$ 
in the non-unitary case.
Since $l = |k+\Delta-m|-m$, the two roots are
\eq
s' = -\frac12 
\left |k+\Delta-m\right | 
\pm \frac12 (k-\Delta+m)
\,.
\en
We also need $s'$ to satisfy the inequalities dictated by 
the decomposition (\ref{Vsfusion}),
\eq
-|s| \le s'
\,,\qquad
-l+|s| \le s'
\,,\qquad
s' \le |s|
\,.
\en
For both the annihilation modes with $k+\Delta-m \ge m$ and the 
creation modes with $k+\Delta-m \le -m$,
we calculate the two roots $s'$, then check for which of the two values of 
$\Delta$ and for which values of $k$ each root satisfies all the 
inequalities (assuming $d>2$ so that $m>0$):
\vskip2ex

\begin{center}
\begin{tabular}{l|l|ll}
& \multicolumn{1}{c|}{root} & \multicolumn{2}{c}{solutions}
\\
\hline
$k+\Delta-m \ge m$  & $s' =-k$         & 
\multicolumn{2}{c}{none}
\\
                  & $s' =m-\Delta$   & $\Delta = m+|s|\,,$ & $k\ge \Delta$
\\[1ex]
$k+\Delta-m \le -m$ & $s' =-m+\Delta$  & $\Delta = m+|s|\,,$ &$ k\le -\Delta$
\\
                  & $s' =k $         & $\Delta = -|s|\,,$  & 
                  $\Delta\le k \le -\Delta$
\end{tabular}
\end{center}

In the unitary case $\Delta = m+|s|$, we
see that the modes $\phi_{k}$ with
$m-|s|\le k<\Delta$ have been eliminated, because their Casimirs do 
not occur in the decomposition of the tensor product.
The mode expansion in the unitary case is
\be
\label{eq:modeexpansion2}
\phi(x) = \sum_{k+\Delta\in \Integers,\;|k|\ge \Delta} r^{-k-\Delta} 
Y_{l}(\hat x)
\phi_{k}\ .
\ee
where
\eq
\Delta = m +|s|
\,,\qquad
l =\left |k+\Delta-m\right|-m 
= 
\left \{
\begin{array}{ll}
k-\Delta+2|s|\,, & k\ge \Delta
\\[1ex]
-k-\Delta\,, & k\le -\Delta
\,.
\end{array}
\right .
\en
and the mode $\phi_{k}$ is in the irreducible representation
\eq\label{phikrepeven}
\phi_{k} \in V_{(|k|-\Delta+|s|,|s|,\ldots,|s|, -\epsilon s)}
\qquad
\epsilon = \sgn(k)\,.
\en
At this point, it remains possible that some of these modes 
are identically zero, since we have only enforced the radial 
component of the first order differential equation.
When we calculate the singular operator product expansion, we will 
find that all these modes are non-zero.

For the non-unitary case, $\Delta=-|s|$, only a finite number of  
modes can be non-zero,
\be
\phi(x) = \sum_{k-|s|\in \Integers,\,|k|\le |s|} r^{-k+|s|} 
Y_{|s|-k}(\hat x)
\phi_{k}\ .
\ee
\eq
\phi_{k} \in
V_{(|s|,|s|,\ldots,|s|,\epsilon' k)}
\,,\qquad
\epsilon' = \sgn(s)\,.
\en
Note that in the non-unitary case, the
expansion in $r$ is a polynomial. We will
return to the consequences of this observation
in the next section.

Finally, let us discuss the modes in odd dimension $d$.
In this case, there is only one non-trivial Cauchy
representation, $V_{(s,\ldots,s)}$ with $s=1/2$.
The decomposition (\ref{eq:decompdodd}) is
\eq
S_{l}\otimes V_{(s,\ldots,s)}= \mathop\oplus_{s'= \pm 1/2}
V_{(l+s',s,\ldots,s)}
\,,\qquad
s=\frac12
\,,\qquad
l\ge 1
\ .
\en
The quadratic Casimir invariants of the components are
\eq
C_2^{d}(V_{(l+s',s,\ldots,s)}) = C_2^{d}(V_{(s,\ldots,s)})
+ \frac12 \left [(l+s')(l+s'+d-2) - s(s+d-2)
\right ]
\,.
\en
Matching to the Casimirs of the $\phi_{k}$, equation (\ref{eq:Casimirphin}),
gives the condition
\eq
(l+s')(l+s'+d-2) - s(s+d-2) = -\Delta^{2}+k^{2}
\,,
\en
which can be re-written, since $s'=\pm\frac12$,
\eq
\label{eq:doddmatching}
s' |k+\Delta-m| = (m-\Delta) (k+\Delta-m) +m\left (m+\frac12 
-\Delta\right)
\,.
\en
For the unitary case, $\Delta = \frac12 (d-1) = m+\frac12$,
this is
\eq
s' \left |k+\frac12 \right | = -\frac12 \left (k+\frac12 \right) 
\,,
\en
The conformal dimension $\Delta = m+\frac12 $ is an integer, so the 
weights $k$ of the modes are integers,
so the solutions are : $s'=\frac12$ for $k \le -1 $
and $s'=-\frac12$ for $k \ge 0$, $l>0$.

The mode expansion for $d$ odd, in the unitary case,
is obtained by combining with the results of the harmonic expansion, equation 
(\ref{eq:modeexpansion}),
\be
\label{eq:modeexpansion2dodd}
\phi(x) = \sum_{|k|\ge \Delta} r^{-k-\Delta} 
Y_{l}(\hat x)
\phi_{k}\,,
\ee
\eq
\Delta = m +\frac12
\,,\qquad
l =\left |k+\Delta-m\right|-m 
= 
\left \{
\begin{array}{ll}
k-\Delta+1\,, & k\ge \Delta
\\[1ex]
-k-\Delta\,, & k\le -\Delta
\,,
\end{array}
\right .
\en
\eq\label{phikrepodd}
\phi_{k} \in V_{(l-\epsilon s,s,\ldots,s)}
= 
V_{(|k|-\Delta+ s,s,\ldots,s)}
\,,
\qquad
\epsilon = \sgn(k)\,.
\en

For the non-unitary case, $\Delta = -\frac12$,
the matching equation (\ref{eq:doddmatching})
is
\eq
s' \left |k-\frac12-m\right| = \left (m+\frac12\right)k -\frac14
\,.
\en
The only solutions are $k=\frac12, s'=\frac12$ with $l=0$ 
and $k=-\frac12,s'=-\frac12$ with $l=1$.
The mode expansion is thus
\be
\phi(x) = Y_0(\hat x) \phi_{1/2} + r^{1} Y_1(\hat x) \phi_{-1/2}
\ee
\eq
\phi_{\pm1/2} \in
V_{(s,s,\ldots,s)}\ .
\en

\subsection{Summary of the mode expansions}

We have established that the non-trivial unitary Cauchy fields 
are the fields with spin $V_{s}$,
\aeq{
\text{for $d$ odd:} \qquad& s=\frac12\,, \\[1ex]
\text{for $d=2n$ even:} \qquad & s\in\left\{\pm\frac12,\, \pm 1,\, 
\pm\frac32,\, \ldots
\right\}\,, 
}
and scaling dimension
\eq
\Delta =|s|+m\,,\qquad m=\frac12 (d-2)\,,
\en
and with mode expansion
\eq
\phi(x)
= \sum_{|k|\ge \Delta} r^{-k-\Delta}  Y_{|k+\Delta-m|-m}(\hat x) 
\,\phi_{k}\ .
\en
\ateq{3}{
\label{phikrepodd2}
\text{for $d$ odd:}& \qquad& \phi_{k}& \in V_{(|k+s|-m-\epsilon s,s,\ldots,s)}
\qquad
&\epsilon &= \sgn(k)\,, \\[1ex]
\label{phikrepeven2}
\text{for $d=2n$ even:}& \qquad & \phi_{k}& \in V_{(|k|-\Delta+|s|,|s|,\ldots,|s|, -\epsilon s)}
\qquad
&\epsilon &= \sgn(k)\,.
}

\section{Operator products, commutators, and stress-energy 
tensor}

Using the mode expansions, we show now that the commutators of the 
modes of unitary Cauchy 
fields are multiples of the identity operator.
Therefore the correlation functions can be calculated by Wick 
contractions, so the unitary Cauchy fields are free fields.

In a first step we will show that the commutator of two Cauchy fields is completely 
determined by the singular part of the operator product expansion of 
the two fields.
Then we show that the singular part of the operator product expansion 
of two Cauchy fields contains only the identity operator.

We also show that the operator product of two Cauchy fields contains 
a field with the spin and scaling dimension of the stress-energy 
tensor iff $d$ is even and $s=\pm 1/2$ or $s=\pm1$ or $d$ is odd and 
$s=1/2$.  So the only Cauchy fields with 
local stress-energy tensors are the massless spinor fields in any
dimension and the 
free (anti-)self-dual $n$-form fields in even dimensions $d=2n$.

\subsection{Commutators and the singular part of the OPE}\label{ss:sing}
The first step is to show that the commutator of a mode $\phi_{k}$ of a Cauchy 
field $\phi(x)$ with any  field $\psi(x)$
is completely determined by the singular part of the operator product 
expansion of the two fields.

Let us extract the mode $\phi_k$ of the Cauchy field $\phi(x)$ by smearing
over the sphere of radius $r$
with the appropriate vector-valued spherical harmonic,
\be\label{getmode}
\int\limits_{S^{d-1}_r} d\Omega \, F_k(\hat x) \phi(x) = r^{-k-\Delta }\phi_{k}
\,.
\ee
Suppose $\psi(y)$ is any local field.
Let $R(\phi(x)\psi(y))$ be the radially ordered
operator product. 
The commutator of $\phi_k$ with $\psi(y)$ 
can be calculated by
evaluating (\ref{getmode}) at radius $r=|y|+\epsilon$ and
at radius $r=|y|-\epsilon$,
\aeq{
\label{comm}
\int\limits_{S^{d-1}_{|y|+\epsilon}}
-\int\limits_{S^{d-1}_{|y|-\epsilon}} d\Omega \, F_k(\hat x) R(\phi(x)\psi(y))
&=(|y|+\epsilon)^{-k-\Delta }\phi_{k} \psi(y)
- (|y|-\epsilon)^{-k-\Delta }\psi(y)\phi_{k}
\nonumber\\
&=|y|^{-k-\Delta }  \left[ \phi_{k} ,\psi(y)\right] + O(\epsilon)
\,.
}
If the OPE of $\phi(x)\psi(y)$ is non-singular, then $R(\phi(x)\psi(y) )$ is bounded
in the
integrals on the left-hand side, 
so that the left-hand side is $O(\epsilon)$.
Sending $\epsilon\rightarrow0$, we see that
the commutator $\left[ \phi_{k} ,\psi(y)\right]$ vanishes.
Therefore the commutator only depends on
the singular part of the OPE.

In fact, the deformability property of Cauchy fields
as described in section \ref{sect:deformability}
will allow us to deform the integration domain on the lhs of (\ref{comm})
to a small sphere centered at $y$.  We then see that the commutator 
depends only on the singular part of the OPE of $\phi(x)\psi(y)$ that 
is at least as singular as $|x-y|^{-(d-1)}$.
For the present purpose we do not need this stronger result.

\subsection{Commutators of Cauchy fields}
The analysis in section~\ref{s:modes}
showed  that there are no modes $\phi_{k}$ in the range $-\Delta<k<\Delta$.
This has drastic consequences for the
operator product expansions of Cauchy fields.

Suppose $\phi_{1}(x)$ and $\phi_{2}(x)$ are Cauchy fields.
We can suppose, without loss of generality,
that $\Delta_{1}\ge\Delta_{2}$ (exchanging 
$\phi_{1}\leftrightarrow\phi_{2}$ if necessary).

The conformal highest weight state corresponding to $\phi_{2}(x)$ is
\eq
\phi_{2}(0) \ket0 = \phi_{2,-\Delta_{2}} \ket0
\en
The operator product $\phi_{1}(x) \phi_{2}(0)$ is given by
\eq\label{CCOPE}
\phi_{1}(x)\,\phi_{2}(0) \ket0 
= \sum_{|k|\ge \Delta_{1}} r^{-k-\Delta_{1}}  Y_{|k+\Delta_{1}-m|-m}(\hat x) 
\,\phi_{1,k}\, \phi_{2,-\Delta_{2}}\ket0 
\,.
\en
The singular part is
\eq
\left(\phi_{1}(x)\,\phi_{2}(0)\right)_{\mathit{sing}} \ket0 = 
\sum_{k\ge\Delta_{1}}r^{-k-\Delta_{1}} Y_{k+\Delta_{1}-2m}(\hat x)  
\,\phi_{1,k}\, \phi_{2,-\Delta_{2}}\ket0 
\,.
\en
The state
\eq
\ket{\phi'_{k-\Delta_{2}}} = \phi_{1,k}\,\phi_{2,-\Delta_{2}}\ket0
\en
has conformal weight 
$-k+\Delta_{2}$.
But $k\ge\Delta_{1}\ge\Delta_{2}$.
By unitarity, there are no states 
with conformal weight $<0$.
Therefore
$\ket{\phi'_{k-\Delta_{2}}} =0$
unless $k=\Delta_{1}=\Delta_{2}$,
in which case $\ket{\phi'_{k-\Delta_{2}}}$
has weight 0, so must be proportional to the ground state,
\eq
\phi_{1,k}\,\phi_{2,-\Delta_{2}}\ket0  = \delta_{k,\Delta_{1}} C_{12} \ket0
\,,\qquad \Delta_{1}=\Delta_{2}
\,.
\en
Therefore, using $\Delta_{1}=|s_{1}|+m$,
\eq
\left(\phi_{1}(x)\,\phi_{2}(0)\right)_{\mathit{sing}} =
\left\{
\begin{array}{ll}
0\,, & \Delta_{1}\ne \Delta_{2}\,,
\\[1ex]
r^{-2\Delta_{1}} Y_{2|s|_{1}}(\hat x)  \, C_{12} \mathbf{1}
\,,\quad
& \Delta_{1}= \Delta_{2}\,.
\end{array}
\right .
\en
From the argument in section~\ref{ss:sing} it thus
follows that all commutators between Cauchy fields
of different scaling dimensions must vanish,
and all commutators of Cauchy fields of the same scaling dimension must be 
proportional to the identity operator.

We have a
decomposition of any Cauchy field $\phi(x)$ into creation and annihilation operators,
so we can evaluate any
correlation function by applying Wick's theorem, commuting 
destruction operators to the right and creation operators to the left.
It thus follows that all correlation functions of unitary Cauchy fields 
factorize into two point functions.
All unitary Cauchy fields are thus indeed free fields.

\subsection{Existence of stress-energy tensor}

Suppose $\phi(x)$ is a Cauchy field of dimension $\Delta$.
If there is a local energy momentum tensor $T_{\mu\nu}(x)$, it must 
have dimension $d$.  The operator product 
$T_{\mu\nu}(x)\; \phi(0)$ must contain the field $\phi(0)$ with a 
canonical non-zero coefficient.  Therefore
\eq
\expval{\phi^{\dagger}(\infty) \,T_{\mu\nu}(x)\,\phi(0)} \ne 0\,.
\en
Therefore, by a global conformal transformation
\eq
\expval{T_{\mu\nu}(\infty)\,\phi^{\dagger}(x) \,\phi(0)} \ne 0\,.
\en
Therefore the operator product of $\phi^{\dagger}(x)\,\phi(0)$ must contain a 
dimension $d$ field.
But the lowest dimension field occurring in 
the operator product $\phi^{\dagger}(x)\,\phi(0)$, besides the identity,
has dimension $2\Delta$.
So there cannot be a local energy momentum tensor $T_{\mu\nu}(x)$
unless $2\Delta \le d$.  Since $\Delta=|s|+(d-2)/2$, this is $|s|\le 
1$.

So a Cauchy field with $|s|>1$ cannot occur in a conformal field 
theory with a local energy-momentum tensor.
This is a manifestation of the Weinberg-Witten theorem \cite{Weinberg:1980kq}
which says that massless fields with spin $j>1$ cannot
couple to a local stress-energy tensor.

\section{More on modes}

Now that we know that every unitary Cauchy conformal field is free, 
we can finish the analysis of the mode expansion.

\subsection{The modes $\phi_{k}$, $|k|\ge \Delta$ are all non-zero}\label{ss:nonzero}

Suppose $\phi(x)$ is a unitary Cauchy field of spin $V_{s}$
and scaling dimension $\Delta = |s|+n-1$.
We show that all the modes 
$\phi_{k}$, $|k|\ge \Delta$ are non-zero
by calculating the commutators of the modes recursively using 
the raising operators $P_{\mu}$, 
and seeing that the commutators are non-zero.

Let $\phi^{\dagger}(x)$ be the adjoint field, so $\expval{\phi^{\dagger}(x)\,\phi(0)}\ne
0$.  The scaling dimension of $\phi^{\dagger}(x)$ is also $\Delta$.  
The spin of $\phi^{\dagger}(x)$ must be one of $V_{\pm s}$.  Which one will be determined later.

Write the mode expansions (\ref{eq:modeexpansion2}) in the form
\eq
\label{eq:modeexpansion3}
\phi(x) = \sum_{k+\Delta\in \Integers,\;|k|\ge \Delta} r^{-k-\Delta} 
\phi_{k}(\hat x )
\,,\qquad
\phi^{\dagger}(x) = \sum_{k+\Delta\in \Integers,\;|k|\ge \Delta} r^{-k-\Delta} 
\phi^{\dagger}_{k}(\hat x )
\,.
\en
The (anti-)commutators of the modes are multiples of the identity,
so
\eq
\label{eq:(anti)commutators1}
[\phi^{\dagger}_{k}(\hat x )\,,\phi_{k}(\hat x )]_{\mp}
= C_{k}(\hat x) \mathbf1
\,.
\en
The commutator is used for $s\in \Integers$, the anti-commutator for $s\in 
\frac12+\Integers$.

The commutators of $P_{\mu}$ with the modes are obtained by 
calculating
\aeq{
[x^{\mu} P_{\mu},\, \phi(x)] & = x^{\mu} \partial_{\mu}\phi(x) \\[1ex]
\sum_{k+\Delta\in \Integers,\;|k|\ge \Delta} r^{-k-\Delta} 
x^{\mu}[ P_{\mu},\, \phi_{k}(\hat x)]
&= 
\sum_{k+\Delta\in \Integers,\;|k|\ge \Delta} r^{-k-\Delta} 
(-k-\Delta)
\phi_{k}(\hat x)
}
which gives
\eq
[ \hat x^{\mu} P_{\mu},\, \phi_{k+1}(\hat x)]
= 
(-k-\Delta)
\phi_{k}(\hat x)
\,.
\en
Then we derive the recursion relation:
\aeq{
0 & = \left [\hat x^{\mu} P_{\mu},\,[\phi^\dagger_{k+1}(\hat x) 
,\,\phi_{-k}(\hat x)]_{\mp}\right ]
\\
& = \left [[\hat x^{\mu} P_{\mu},\,\phi^\dagger_{k+1}(\hat x) ],\,\phi_{-k}(\hat x)\right ]_{\mp}
+
\left [\phi^\dagger_{k+1}(\hat x) ,\,[\hat x^{\mu} P_{\mu},\,\phi_{-k}(\hat x)]\right ]_{\mp}
\\
& = \left [(-k-\Delta)\phi^\dagger_{k}(\hat x) ,\,\phi_{-k}(\hat x)\right ]_{\mp}
+
\left [\phi^\dagger_{k+1}(\hat x) ,\,(k+1-\Delta)\phi_{-k-1}(\hat x)\right ]_{\mp}
}
which is
\eq
(k+1-\Delta) C_{k+1}(\hat x)
=
(k+\Delta)C_{k}(\hat x)
\en
which we solve to get
\ateq{2}{
\label{eq:recursion1}
C_{k}(\hat x)
&=
\binom{k+\Delta-1}{2\Delta-1}
C_{\Delta}(\hat x)
\qquad & k&\ge \Delta
\\[1ex]
C_{k}(\hat x)
&=
\binom{-k+\Delta-1}{2\Delta-1}
C_{-\Delta}(\hat x)
\qquad & k&\le -\Delta
\,.
}
The second equation is equivalent to the first under 
$\phi\leftrightarrow\phi^{\dagger}$,
so we only need to consider $k\ge \Delta$.

The mode $\phi_{-\Delta}$ has $l=0$, so $\phi_{-\Delta}(\hat x)$ is 
independent of $\hat x$, so the two-point function is
\eq
\expval{\phi^{\dagger}(x) \phi(0)} 
= \bra0 |x|^{-2\Delta}\phi^{\dagger}_{\Delta}(\hat x)
\, \phi_{-\Delta}(\hat x) \ket0
= |x|^{-2\Delta} C_{\Delta}(\hat x)
\en
therefore $C_{\Delta}(\hat x)$ is not identically zero, therefore,
by the recursion relation (\ref{eq:recursion1}),
none of the $C_{k}(\hat x)$, $k\ge \Delta$ are identically zero,
therefore none of the modes $\phi_{k}$, $\phi^{\dagger}_{-k}$, $k\ge 
\Delta$, are identically 
zero.  The same is true exchanging 
$\phi\leftrightarrow\phi^{\dagger}$.
So all the modes $\phi_{k}$, $\phi^{\dagger}_{k}$, $|k|\ge 
\Delta$, are non-zero.

\subsection{The spin of $\phi^{\dagger}(x)$}
The two point function is
\eq
\expval{\phi^{\dagger}(x)\,\phi(0)} =
\bra0 \phi^{\dagger}_{\Delta} \, \phi_{-\Delta} \ket 0
\en
with
\eq
\phi_{-\Delta}\in V_{s}
\,,
\en
so  $\phi^{\dagger}_{\Delta}$ must be in $V_{s}^{*}$, the dual space 
to $V_{s}$.
The representation $V_{s}$ is unitary, 
$
V_{s} = V_{s}^{\dagger} = \bar V_{s}^{*}
$,
so the dual space is the same 
as the complex conjugate,
$
V_{s}^{*} = \bar V_{s}
$.
For $d=2n$ even,
all of the $V_{s}$ are generated by tensor products of $V_{\pm\frac12}$ 
with itself, and $V_{\pm\frac12}$ is the chiral spinor representation.
For $d$ odd, $V_{\frac12}$ is the spinor representation.
So, from the reality properties of spinors,
\eq
\phi^{\dagger}_{\Delta}\in V_{s}^{*} = \bar V_{s} = 
\left \{
\begin{array}{ll}
V_{s} &  d=2n,\; n \text{ even} \\
V_{-s} &  d=2n,\; n \text{ odd} \\
V_{s} &  d \text{ odd} \,.
\end{array}
\right .
\en
Comparing to
the representation of $\phi^{\dagger}_{\Delta}$ as given
by  (\ref{phikrepodd2}) and (\ref{phikrepeven2}),
we find that the spin of the adjoint field $\phi^{\dagger}(x)$ must be
\eq
V_{\bar s} = 
\left \{
\begin{array}{ll}
V_{-s} &  d=2n,\; n \text{ even} \\
V_{s} &  d=2n,\; n \text{ odd} \\
V_{s} &  d \text{ odd} \,.
\end{array}
\right .
\en

\subsection{Invariant (anti-)commutators of the modes}
The mode $\phi_{-k}$
lies in a representation
given by  (\ref{phikrepodd2}) or (\ref{phikrepeven2}),
and the mode $\phi_{k}^\dagger$ lies in the
dual representation.
Using upper indices $\alpha,\,\beta$ for the representation of
$\phi_{-k}$, and lower indices $\alpha,\,\beta$ for the dual 
representation,
the (anti-)commutator of the modes takes the form
\eq
\label{eq:ck}
[\phi^{\dagger}_{k,\beta},\,\phi_{-k}^{\alpha}]_{\mp} =  
c_{k}\delta^{\alpha}_{\beta}
\en
where $c_{k}$ is a number.

The terms in the mode expansions (\ref{eq:modeexpansion3})
take the form
\eq
\phi_{-k}(\hat x) = Y_{l}(s,-k;\hat x)_{\alpha} \phi^{\alpha}_{-k}
\,,\qquad
\phi^{\dagger}_{k}(\hat x) = Y_{l}(\bar s,k;\hat x)^{\beta}\phi^{\dagger}_{k,\beta}
\,,
\en
where $Y_{l}(s,-k;\hat x)_{\alpha}$ is the vector spherical harmonic 
expressing the Clebsch-Gordan coefficients between $S_{l}$, $V_{s}$, 
and the representation of $\phi_{-k}$, only the last of which is 
labeled by an explicit index, $\alpha$,
and similarly for $Y_{l}(\bar s,k;\hat x)^{\beta}$.

Equation (\ref{eq:(anti)commutators1}) becomes
\eq
C_{k}(\hat x) = c_{k} Y_{l}(\bar s,k;\hat x)^{\alpha} Y_{l}(s,-k;\hat x)_{\alpha}
\,.
\en
In particular, the starting point of the recursion formula is
\eq
C_{\Delta}(\hat x) = c_{\Delta} Y_{2|s|}(\bar s,\Delta;\hat 
x)^{\alpha} Y_{0}(s,-\Delta;\hat x)_{\alpha}
= c_{\Delta} Y_{2|s|}(\bar s,\Delta;\hat x)
\,,
\en
where $Y_{2|s|}(\bar s,\Delta;\hat x)$ is the Clebsch-Gordan for 
$S_{2|s|}$, $V_{\bar s}$, and $V_{s}$.
The 2-point function is
\eq
\expval{\phi^{\dagger}(x) \phi(0)} 
= \bra0 |x|^{-2\Delta}\phi^{\dagger}_{\Delta}(\hat x)
\, \phi_{-\Delta}(\hat x) \ket0
= |x|^{-2\Delta} C_{\Delta}(\hat x)
= |x|^{-2\Delta} c_{\Delta} Y_{2|s|}(\bar s,\Delta;\hat x)
\,.
\en
The result (\ref{eq:recursion1}) of the recursion
gives us
\eq
c_{k} Y_{l}(\bar s,k;\hat x)^{\alpha} Y_{l}(s,-k;\hat x)_{\alpha}
=
\binom{k+\Delta-1}{2\Delta-1}
c_{\Delta} Y_{2|s|}(\bar s,\Delta;\hat x)
\,,\qquad  k\ge \Delta
\,,
\en
which determines the numbers $c_{k}$,
after some group-theoretic work which we 
refrain from doing.

\section{Deformability}\label{sect:deformability}
\subsection{Smearing currents}
Suppose $\phi(x)$ is a Cauchy field in the Cauchy conformal 
representaton $(V,\Delta)$, not necessarily unitary.
We want to be able to use the same deformation of contour arguments
for $\phi(x)$ as for a holomorphic field in 2 dimensions.
In particular, we would like to calculate
a commutator $[\phi_{k},\,\psi(y)]$ by deforming the integrals for the 
mode $\phi_{k}$ to an integral over a small sphere centered at $y$,
so that the commutator can be extracted from the operator product 
expansion of $\phi(x)\,\psi(y)$.

A deformable integral over codimension 1 surfaces in space-time 
is given by a conserved current.
So we want to represent each mode of $\phi(x)$ by smearing over a codimension 1 surface $S$
by a vector-valued current $f^\mu(x)$,
by
\eq\label{fmode}
\phi[f,S] = \int_{S} d^{d-1}x \;\hat n_{\mu} \expval{f^{\mu}(x),\, \phi(x)}
\,,\qquad
f^{\mu}(x) \in V
\en
where the inner-product on the rhs is the invariant hermitian 
inner-product on $V$.
We can deform $S$ to any cobounding $S'$
by Stokes' theorem, if
\eq\label{nodiv}
\partial_{\mu}  \expval{f^{\mu}(x),\, \phi(x)}= 0
\,.
\en
Since $\phi(x)$ satisfies a linear first order differential equation that  
contains only a linear combination of derivatives of $\phi$,
$f^{\mu}(x)$ must separately satisfy
\be
\label{eq:fconserved}
\partial_\mu f^\mu(x) = 0
\ee
and
\be\label{fcond2}
\expval{f^{\mu}(x),\, \partial_{\mu}\phi(x)} = 0\ .
\ee
The Cauchy differential equation (\ref{eq:Cauchydiffeqn}) can be written
\eq
\partial_{\nu}\phi = (1-\hat P)^{\mu}_{\nu}\partial_{\mu}\phi
\en
where $\hat P$ is the self-adjoint projection on the null subspace in level 
1 of the Verma module.
So equation (\ref{fcond2}) is equivalent to
\eq
\label{fcondp}
 (1-\hat P)^{\mu}_{\nu}  f^{\nu}(x) = 0
\,.
\en
The smearing currents $f^{\mu}(x)$ satisfying  (\ref{eq:fconserved}) and (\ref{fcondp}) 
are exactly those
defining a mode $\phi[f,S]$ that can be evaluated on any deformation 
$S'$
of $S$.

\subsection{Enough smearing currents}
We want to show now 
that there are enough smearing currents
satisfying
(\ref{eq:fconserved}) and (\ref{fcondp})
to capture all the modes of $\phi(x)$.
This means that the contour deformation technique can be used to 
calculate with the modes. 

For simplicity, we take the codimension 1 surface $S$ to be $\Reals^{d-1}$.
The Cauchy differential equation is first order,
so it will certainly be possible to covariantize.
The complexified tangent space of space-time is $\Complexes^{d}$.  
Decompose it into the tangential and normal subspaces,
\eq
\Complexes^{d} = \Complexes^{d-1}\oplus \Complexes
\,,\qquad
\partial_{\mu} = (\vec\partial, \partial_{d})\,.
\en
The projection matrix 
\eq
\hat P : \Complexes^{d}\otimes V \rightarrow 
\Complexes^{d}\otimes V
\en
that enters
in the Cauchy first order differential equation (\ref{eq:Cauchydiffeqn}) 
and in equation (\ref{fcondp}) for the smearing current
decomposes into a block matrix
\eq
\label{eq:Pblock}
\hat P =
\left (
\begin{array}{c|c}
	\overleftrightsmallarrow{P} & \vec P
	\\\hline
	\raisebox{-4pt}{$\vec P^{\dagger}$}  &  \raisebox{-4pt}{$\hat 
		P^{d}_{d}$}
\end{array}
\right )
\,.
\en
$\hat P$ is a self-adjoint projector, so
the operators $\hat P^{d}_{d}$ and $\overleftrightsmallarrow{P}$ are self-adjoint,
but by themselves they are not projectors.
The projector condition $\hat P^{2}=\hat P$ is
\ateq{2}{
	\overleftrightsmallarrow{P}^{2}+\vec P \vec P^{\dagger} 
	&= \overleftrightsmallarrow{P}
	\;\,\text{ or } &
	\vec P \vec P^{\dagger}  &= 
	\overleftrightsmallarrow{P}(1-\overleftrightsmallarrow{P})\,,
	\label{eq:Phatproj4}
	\\
	\label{eq:Phatproj3}
	\overleftrightsmallarrow{P}\vec P+\vec P \hat P^{d}_{d}
	&= \vec P
	\;\;\;\text{ or }&
	\vec P \hat P^{d}_{d}
	&= (1-\overleftrightsmallarrow{P})\vec P
	\,,
	\\
	\vec P^{\dagger} \overleftrightsmallarrow{P}+\hat P^{d}_{d} \vec P^{\dagger}
	&= \vec P^{\dagger}
	\;\;\text{ or }\;&
	\vec P^{\dagger} \overleftrightsmallarrow{P}
	&= (1-\hat P^{d}_{d})\vec P^{\dagger}
	\,,
	\label{eq:Phatproj2}
	\\
	\vec P^{\dagger} \vec P +(\hat P^{d}_{d})^{2}& = \hat P^{d}_{d}
	\;\;\text{ or } &
	\vec P^{\dagger} \vec P & = \hat P^{d}_{d}(1-\hat P^{d}_{d})
	\label{eq:Phatproj1}
	\,.
}
Decomposing $f$ into components $(\vec{f},f_d)$ and
using the block matrix decomposition (\ref{eq:Pblock}) of $\hat P$,
equation (\ref{fcondp}) becomes
\aeq{\label{eq:feqn1}
\overleftrightsmallarrow{P} \vec f(x)+\vec P f_{d}(x)  &= \vec f(x)
\\
\label{eq:feqn2}
\vec P^{\dagger}\vec f(x) + \hat P^{d}_{d} f_{d}(x) &= f_{d}(x)
\,.
}
Our approach is to pick $f_d(x)$, and then use 
the first equation (\ref{eq:feqn1}) to determine $\vec{f}$,
solving
\eq
\left(1-\overleftrightsmallarrow{P}\right)\vec f(x)
= 
\vec P f_{d}(x) 
\,,
\en
to express $\vec f$ in terms of $f_{d}$.
This has a solution because the projector condition (\ref{eq:Phatproj4}) 
implies that the image of $\vec P$ is orthogonal to the eigenspace 
$\overleftrightsmallarrow{P}=1$.
We can thus define the inverse of $1-\overleftrightsmallarrow{P}$ to be zero on 
the eigenspace $\overleftrightsmallarrow{P}=1$
to obtain
\eq
\label{eq:fveceqn}
\vec f(x)
= 
\big(1-\overleftrightsmallarrow{P}\big)^{-1}\vec P f_{d}(x) 
+ \vec f(x)_{1}
\,,\qquad
\overleftrightsmallarrow{P} \vec f(x)_{1} = \vec f(x)_{1} 
\,,
\en
for some arbitrary $\vec f(x)_{1}$ in the eigenspace $\overleftrightsmallarrow{P}=1$.
The spatial piece $\vec f$ of the current is thus determined by $f_d$ up to the component
of eigenvalue 1 under $\overleftrightsmallarrow{P}$.
Using the projector conditions (\ref{eq:Phatproj4}-\ref{eq:Phatproj1}),
it is straightforward to check that (\ref{eq:feqn2}) is then automatically satisfied.
This shows that (\ref{fcondp}) imposes no further constraints on 
our choice of $f_d$.
The only remaining condition on $f^{\mu}(x)$ is (\ref{eq:fconserved}),
\eq
\label{eq:fconserved2}
\partial_{d}f_{d}(x) = -\vec \partial \cdot \vec f(x)
\,.
\en
It follows that $\partial_d f_d$ as given by  (\ref{eq:fconserved2})
and $\partial_d\vec f$ as given by (\ref{eq:fveceqn})
can be integrated for arbitrary initial data $f_{d}(x)$ on $S$.
The corresponding mode is
\eq
\phi[f,S] = \int_{S} d^{d-1}\vec x \; \expval{f_{d}(x),\,\phi(x)}\ .
\en
Since $f_{d}(x)$ is arbitrary on $S$, there are enough deformable smearing currents to capture every mode of 
$\phi(x)$. In principle, for $V=V_{s}$, we could now construct 
explicit smearing 
functions $f_{k}^{\mu}(x)$ such that
\eq
\phi[f_{k},S^{d-1}] = \phi_{k}
\en
for the modes $\phi_k$ defined in section~\ref{s:modes}, but
we will refrain from doing so here.

\subsection{Duality}\label{s:overdet}
Equation (\ref{fmode}) suggests a
duality between the Cauchy fields $\phi(x)$ and the smearing currents 
$f^{\mu}(x)$.
But not every initial data $\phi(x)$ on $S$ can be integrated to a 
solution of the Cauchy equation.
We will see below that the Cauchy equation is over-determined.
So there must be an equivalence relation -- a gauge symmetry -- on 
the smearing currents.

In fact,
equations (\ref{eq:fconserved}) and (\ref{fcondp}) on $f^{\mu}(x)$
have  the gauge symmetries
\be
f^\mu(x)\rightarrow f^\mu(x)+ \delta f^{\mu}(x)
\ee
\eq
\label{eq:gaugesymm}
\delta f^{\mu}(x) =  \partial_{\sigma} g^{\mu\sigma}(x)
\,,\qquad
g^{\mu\sigma}(x)=-g^{\sigma\mu}(x)
\,,\qquad
(1-\hat P)^{\mu}_{\nu} g^{\nu\sigma}(x) = 0
\,.
\en
The mode $\phi[f]$ is gauge-invariant because (\ref{eq:gaugesymm}) 
implies
\aeq{
\phi[f+\delta f,S]-\phi[f,S] &= \int_{S} d^{d-1}x \;\hat n_{\mu} 
\expval{\partial_{\sigma}g^{\mu\sigma}(x),\, \phi(x)}
\\
&= \int_{S} d^{d-1}x \;\hat n_{\mu} 
\expval{{g^{\sigma\mu}(x),\,\partial_\sigma} \phi(x)}
\\
&=0\,.
}

We would want to show
that the solutions of equations (\ref{eq:fconserved}) and
(\ref{fcondp}) on $f^{\mu}(x)$ 
modulo the gauge transformations (\ref{eq:gaugesymm})
are exactly dual to solutions of the Cauchy 
differential equation (\ref{eq:Cauchydiffeqn}),
with the pairing given by (\ref{fmode}).
We do not know how to do that.
It requires describing the space of smearing functions $f_{d}(x)$ 
modulo gauge transformations, and the space of initial data $\phi(x)$ 
satisfying the overdetermination conditions.
We suspect that \cite{Wallach:1997} is relevant.

Here, we will only take a first look at the overdetermination 
conditions.
Again taking the codimension 1 surface $S$ to be $\Reals^{d-1}$,
we ask for the conditions
that $\phi(x)$ must satisfy on $S$ in order to extend to a solution of
the Cauchy equation (\ref{eq:Cauchydiffeqn}) off of $S$.
We derive only the first such over-determination condition,
which is a first order differential 
equation on $S$.
Using the block decomposition (\ref{eq:Pblock}) of $\hat P$,
the Cauchy differential equation (\ref{eq:Cauchydiffeqn}) 
becomes
\aeq{
	\label{eq:diffeqn1}
	\overleftrightsmallarrow{P}\vec \partial\phi +\vec P \partial_{d}\phi
	&=0
	\\
	\label{eq:diffeqn2}
	\vec P^{\dagger}\vec \partial\phi +\hat P^{d}_{d} \partial_{d}\phi
	&= 0
	\,.
}
The Cauchy condition is equivalent to the invertibility of $\hat 
P^{d}_{d}$, so (\ref{eq:diffeqn2}) can be written
\eq
\label{eq:normaleqn}
\partial_{d}\phi = - \big (\hat P^{d}_{d}\big )^{-1}
\vec P^{\dagger}\vec \partial\phi
\,,
\en
which gives the normal derivative in terms of the data on $S$.
This is the Cauchy property.

Now we substitute for $\partial_{d}\phi$ in equation 
(\ref{eq:diffeqn1}), getting
\eq
\label{eq:overdetermine}
\overleftrightsmallarrow{P_{1}} \vec \partial\phi = 0
\,,
\en
where
\eq
\overleftrightsmallarrow{P_{1}}
=
\overleftrightsmallarrow{P}
-\vec P \big (\hat P^{d}_{d}\big )^{-1}\vec P^{\dagger}
\,.
\en
From the identities (\ref{eq:Phatproj4})--(\ref{eq:Phatproj1})
it follows that
\eq
\big(\overleftrightsmallarrow{P_{1}}\big)^{2} = \overleftrightsmallarrow{P_{1}} 
\,,\qquad
\overleftrightsmallarrow{P_{1}} \overleftrightsmallarrow{P}
= \overleftrightsmallarrow{P}\overleftrightsmallarrow{P_{1}}
= \overleftrightsmallarrow{P}
\en
so $\overleftrightsmallarrow{P_{1}}$ is the projection on the eigenspace
$\overleftrightsmallarrow{P}=1$.

The first-order differential equation (\ref{eq:overdetermine}) 
is the first over-determination  condition.
It contains no derivatives in the normal direction, so it is a 
differential equation on $S$.
The Cauchy differential equation (\ref{eq:Cauchydiffeqn}) is 
equivalent to the combination of (\ref{eq:normaleqn}) and (\ref{eq:overdetermine}).
In Appendix~\ref{app:overdet} we work out the first-order over-determination
condition explicitly for the unitary Cauchy representations $V=V_s$.

The covariant form of the first-order over-determination condition is the 
differential equation on $S$,
\eq
\hat P_{1}(x)^{\mu}_{\nu}\partial_{\mu}\phi(x) = 0\,,
\en
where $\hat P_{1}(x)$ is the projection on the eigenspace with eigenvalue 
1 of the operator
\eq
\overleftrightsmallarrow{P}(x) =[ P(T_{x}S)\otimes 1]\; \hat P \; 
[P(T_{x}S)\otimes 1]
\en
on $\Complexes^{d}\otimes V$, where $P(T_{x}S)$ is the projection on the complexified tangent space 
to $S$ at $x$, which is a subspace of $\Complexes^{d}$.

The first order over-determination 
condition (\ref{eq:overdetermine}) is a necessary condition on the 
initial data $\phi(x)$ on $S$, but it is not necessarily 
a sufficient condition for integrating the Cauchy equation off $S$.
Equation (\ref{eq:normaleqn}) can be integrated to determine $\phi(x)$ 
uniquely on any nearby surface  $S'$.  But further integration 
requires $\phi(x)$ on $S'$ to continue to satisfy 
(\ref{eq:overdetermine}).  We need the integrability condition
\eq
\label{eq:integrability1}
\partial_{d} \overleftrightsmallarrow{P_{1}} \vec \partial\phi = 0 
\,.
\en
Let us write (\ref{eq:normaleqn}) as
\eq
(\partial_{d} +A_{d})\phi = 0
\,,\qquad 
A_{d}
=A_{d}^{j}\partial_{j}
= - \big (\hat P^{d}_{d}\big )^{-1}
\big(\vec P^{\dagger}\big)^{j}\partial_{i}
\en
and  (\ref{eq:overdetermine}) as
\eq
A_{i} \phi = 0
\,,\qquad 
A_{i} = \overleftrightsmallarrow{P_{1}}^{k}_{i}  \partial_{k}\phi
\,.
\en
The integrability condition (\ref{eq:integrability1}) is
\eq
0 = \partial_{d} A_{i}\phi 
=
A_{i}\partial_{d} \phi 
=
A_{i} (-A_{d}) \phi
= [ A_{d},\,A_{i}]\phi - A_{d} A_{i}\phi
= [ A_{d},\,A_{i}]\phi
\,,
\en
since we already have the first-order condition $A_{i}\phi=0$.
So we need the second-order over-determination condition
\eq
[ A_{d},\,A_{i}]\phi  = [A_{d}^{j},\, A_{i}^{k}] 
\partial_{j}\partial_{k} \phi = 0
\,.
\en
If the second-order condition over-determination condition follows 
from the first-order condition, then we are done.
If not, then we have to impose the second-order condition in addition 
to the first-order condition, and then check the integrability of the 
second-order condition, and so on.

In the unitary case, whatever the complete set of over-determination conditions, we know 
that, when $S$ is the unit sphere, all the solutions are the $\phi_{k}(\hat 
x)$, because these are the only possible modes and all are non-zero.

The structure of the over-determination conditions is determined by 
the pattern of highest weight vectors in the Verma module.
The null space $(P_{d} + A_{d}^{j}P_{j})\phi$ generates a submodule 
of the Verma module -- all null states, all perpendicular to the entire 
Verma module. Complementary to this null submodule is the submodule
with basis
\eq
\left\{P_{i_{1}}\cdots P_{i_{N}}\ket\phi\right\}
\,.
\en
This is the $\so(d-1,2)$ Verma module generated by the $\so(d)$ 
representation $V$ considered as a representation of $\so(d-1)$.
The first-order over-determination condition corresponds to a null 
space $A_{i}^{k}P_{k}\ket\phi$ on level 1.
The second-order condition corresponds to a null space on level 2.
The second-order condition is independent if the corresponding level 
2 null subspace is not contained in the sub-module generated by the level 
1 null space.
So specifying the over-determination conditions is equivalent to 
finding the minimal set of generators for the full null sub-module of the 
$\so(d-1,2)$ Verma module.

\bigskip
\noindent {\bf Acknowledgments}
\medskip

We thank Matt Buican, Thomas Dumitrescu, Simeon Hellerman,
Siddhartha Sahi, Scott Thomas,
and Sasha Zhiboedov for useful
discussions.
Peter Goddard helped us with the history of 2d holomorphic fields.
Siddhartha Sahi pointed us to the proof  in \cite{defosseux2010}
of the branching rule we used in section \ref{s:constrainthemodesradial}.
Nolan Wallach pointed us to the general unitarity bound in 
\cite{MR733809}, and to \cite{Wallach:1997},
and suggested a possible argument to finish the classification of the non-unitary 
Cauchy fields (section \ref{s:nonunitarydabove4}).
Nicolas Boulanger, Dmitry Ponomarev, Evgeny Skvortsov and Massimo Taronna
gave useful explanation about the status of the conformal Coleman-Mandula
theorem.
We are grateful to them for their help.

DF and CAK were supported by the Rutgers
New High Energy Theory Center.
CAK thanks the Aspen Center for Physics and the
Harvard University High Energy Theory Group for hospitality,
where part of this work was completed.
CAK was supported 
in part by National Science Foundation Grant No.~PHYS-1066293,
by the Swiss National Science Foundation through the NCCR SwissMAP,
and  by U.S. DOE Grants No.~DOE-SC0010008 and
DOE-ARRA-SC0003883.



\appendix
\appendixpage
\vskip2ex

\section{The irreducible representations of $\mathbf{so}(d)$}
\subsection{The highest weights representations}

In this appendix we collect
some basic results in the representation theory of $\mathbf{so}(d)$,
taken from \eg \cite{MR1153249}.
The irreducible representations of $\mathbf{so}(d)$ are written 
$V_{\lambda}$, where $\lambda$ is the highest weight.
For $d=2n$, the highest weights of the irreducible representations are
\eq
\lambda = (\lambda_{1},\ldots, \lambda_{n})
\,,\qquad
\lambda_{1} \ge \lambda_{2} \ge \cdots \ge \lambda_{n-1} \ge |\lambda_{n}|
\,.
\en
For $d=2n+1$, the highest weights of the irreducible representations 
\eq
\lambda = (\lambda_{1},\ldots, \lambda_{n})
\,,\qquad
\lambda_{1} \ge \lambda_{2} \ge \cdots \ge \lambda_{n-1} \ge \lambda_{n} \ge 0
\,.
\en
In both cases, the $\lambda_{i}$ are all integers (for the vector representations)
or all half-integers (for the spinor representations).

Some notable representations are given by:
\ateq{3}{
&\text{for all $d$:}\qquad  &&\text{trivial representation}& V_{0} &= V_{(0,\cdots,0)} 
\\[2ex]
&                             &&\text{fundamental representation} & \Complexes^{d} &= V_{(1,0,\ldots,0)}
\\[2ex]
&&&\text{symmetric traceless $l$-tensors}\qquad & S_{l}
&= V_{(l,0,\ldots,0)}
\\[3ex]
&\text{for $d=2n$:}\qquad  &&\text{$p$-forms, $0\le p \le n-1$}
&
\Lambda^{p} &= 
V_{
{\scriptstyle (}{\scriptstyle\underbrace{\scriptstyle 
1,\ldots,1}_{p}},{\underbrace{\scriptstyle 0,\ldots,0}_{n-p}})} 
\\[1ex]
&&&\text{(anti-)self-dual $n$-forms} \qquad &
\Lambda^{n}_{\pm} &= V_{(1,1,\ldots,1,\pm 1)}
\\[2ex]
&&&\text{chiral spinors}
&
S_{\pm} &= V_{\left(\frac12,\frac12, \cdots, \frac12, \pm 
\frac12\right )}
\\[3ex]
&\text{for $d=2n+1$:}\qquad  &&\text{$p$-forms, $0\le p \le n$}
\qquad&
\Lambda^{p} &= 
V_{
{\scriptstyle (}{\scriptstyle\underbrace{\scriptstyle 
1,\ldots,1}_{p}},{\underbrace{\scriptstyle 0,\ldots,0}_{n-p}})} 
\\[1ex]
&&&\text{spinors}
&
S &= V_{\left(\frac12,\frac12, \cdots, \frac12, 
\frac12\right )}
}
The quadratic Casimir invariant --- with the normalization given in equation 
(\ref{eq:Casimir}) --- is, for all $d$,
\eq
\label{eq:quadraticCasimir}
C_{2}^{d}(V_{\lambda}) = \frac1{2}\sum_{i=1}^{n}  \lambda_{i} 
(\lambda_{i}+d-2i)
\en
Decomposing the tensor product $\Complexes^{d}\otimes V_{\lambda}$
into irreducibles gives
\eq\label{sodtensor}
\Complexes^{d}\otimes V_{\lambda} =  
\left \{
\begin{array}{ll}
\mathop\oplus\limits_{\lambda'= \lambda\pm \epsilon_{k}} 
V_{\lambda'}
\qquad & d=2n
\\[2ex]
\mathop\oplus\limits_{\lambda'= \lambda, \lambda\pm \epsilon_{k}} 
V_{\lambda'}
\qquad & d=2n+1
\end{array}
\right .
\en
where $\epsilon_{k}=(0,\ldots,0,1,0,\ldots)$ is the weight that has $1$ in the $k$-th position 
and $0$  elsewhere,
and the sums include all the $\lambda'= \lambda\pm 
\epsilon_{k}$ that are highest weights for $\mathbf{so}(d)$.

Finally we will need 
the branching rules  of an $\mathbf{so}(d)$ irreducible $V_{\lambda}$ decomposing into 
$\mathbf{so}(d-1)$ irreducibles $V^{d-1}_{\mu}$,
\eq
V_{\lambda}  = \mathop\oplus_\mu V^{d-1}_{\mu}
\en
where the sum ranges over all $\mu$ that are highest weights for 
$\mathbf{so}(d-1)$ satisfying $\lambda_{1}-\mu_{1}\in \Integers$ and
\eq
\label{eq:branching2}
\begin{array}{ll}
\lambda_{1} \ge \mu_{1}\ge \lambda_{2} \ge \mu_{2} \ge \cdots \ge 
\lambda_{n-1} \ge \mu_{n-1}\ge |\lambda_{n}|\,,
\qquad & d=2n\,,
\\[3ex]
\lambda_{1}\ge \mu_{1} \ge \lambda_{2} \ge \mu_{2} \ge \cdots \ge \lambda_{n-1}\ge \mu_{n-1} \ge 
\lambda_{n} \ge |\mu_{n}|\,,
\qquad & d=2n+1\,.
\end{array}
\en

\subsection{The eigenvalues of $\hat M$}

In what follows we need to know the eigenvalues $\hat 
M_{\lambda',\lambda}$ of $\hat M$ acting on the
irreducible components $V_{\lambda'}\subset 
\Complexes^{d}\otimes V_{\lambda}$,
\be 
\hat M_{\lambda',\lambda} =  C_{2}^{d}(V_{\lambda}) + C_{2}^{d}(\Complexes^{d}) 
- C_{2}^{d}(V_{\lambda'})\ .
\ee
For the $\lambda'$ of interest,
the ones occurring in (\ref{sodtensor}),
this evaluates to
\bea
\hat M_{\lambda + \epsilon_{k},\lambda} &=& 
k - 1 -\lambda_{k}
\\
\hat M_{\lambda - \epsilon_{k},\lambda} &=& 
\lambda_{k} + d-1-k
\\
\hat M_{\lambda,\lambda} &=&  \frac12 (d-1)
\eea
Recall that the number $\hat M_{\lambda',\lambda}$ occurs as an 
eigenvalue only if $\lambda'$ is actually a highest weight.

To check unitarity, it is important to identify the
largest eigenvalue of $\hat M$.
For $d=2n$, the numbers $\hat M_{\lambda',\lambda}$
satisfy the inequalities\\[1ex]
for $\lambda_{n}>0$:
\eq\label{evenpos}
\hat M_{\lambda + \epsilon_{1},\lambda}
<
\cdots
<
\hat M_{\lambda + \epsilon_{n-1},\lambda}
<
\hat M_{\lambda + \epsilon_{n},\lambda}
<
\hat M_{\lambda - \epsilon_{n},\lambda}
<
\hat M_{\lambda - \epsilon_{n-1},\lambda}
<
\cdots
<
\hat M_{\lambda - \epsilon_{1},\lambda}
\,,
\en
for $\lambda_{n}=0$:
\eq\label{evenzero}
\hat M_{\lambda + \epsilon_{1},\lambda}
<
\cdots
<
\hat M_{\lambda + \epsilon_{n-1},\lambda}
<
\hat M_{\lambda + \epsilon_{n},\lambda}
=
\hat M_{\lambda - \epsilon_{n},\lambda}
<
\hat M_{\lambda - \epsilon_{n-1},\lambda}
<
\cdots
<
\hat M_{\lambda - \epsilon_{1},\lambda}
\,,
\en
for $\lambda_{n}<0$:
\eq\label{evenneg}
\hat M_{\lambda + \epsilon_{1},\lambda}
<
\cdots
<
\hat M_{\lambda + \epsilon_{n-1},\lambda}
<
\hat M_{\lambda - \epsilon_{n},\lambda}
<
\hat M_{\lambda + \epsilon_{n},\lambda}
<
\hat M_{\lambda - \epsilon_{n-1},\lambda}
<
\cdots
<
\hat M_{\lambda - \epsilon_{1},\lambda}
\,.
\en
For $d=2n+1$, the $\hat M_{\lambda',\lambda}$ satisfy\\[1ex]
for $\lambda_{n}>0$:
\eq\label{oddpos}
\hat M_{\lambda + \epsilon_{1},\lambda}
<
\cdots
<
\hat M_{\lambda + \epsilon_{n},\lambda}
<
\hat M_{\lambda,\lambda}
<
\hat M_{\lambda - \epsilon_{n},\lambda}
<
\cdots
<
\hat M_{\lambda - \epsilon_{1},\lambda}
\,,
\en
for $\lambda_{n}=0$:
\eq\label{oddzero}
\hat M_{\lambda + \epsilon_{1},\lambda}
<
\cdots
<
\hat M_{\lambda + \epsilon_{n},\lambda}
<
\hat M_{\lambda,\lambda}
=
\hat M_{\lambda - \epsilon_{n},\lambda}
<
\cdots
<
\hat M_{\lambda - \epsilon_{1},\lambda}
\,.
\en

\subsection{Example: $d=4$ }
Let us give the more familiar expressions for $d=4$ here.
The general formula becomes
\eq
C_{2}^{d}(V_{\lambda}) = \frac1{2} \lambda_{1}(\lambda_{1}+2)
+\frac12 \lambda_{2}^{2}\ .
\en
We usually write $\mathbf{so}(4) = \mathbf{su}(2)_{L} 
\times \mathbf{su}(2)_{R}$.
The irreducible representations of $\mathbf{so}(4)$ are the tensor 
products of $\mathbf{su}(2)_{L,R}$ irreducibles $j_{L},\,j_{R}$.
The corresponding $\mathbf{so}(4)$ highest weight is
\eq
\lambda_{1} = j_{L}+j_{R}
\,,\qquad
\lambda_{2} = j_{L}-j_{R}
\,,
\en
giving
\eq
C_{2}^{d}(V_{\lambda}) = j_{L}(j_{L}+1) + j_{R}(j_{R}+1)
\,.
\en

\section{Classification of Cauchy fields}

Next, we want to classify the Cauchy conformal fields.  That is,  we 
want to determine which representations $V_{\lambda}$ and which 
scaling dimensions $\Delta$ satisfy the algebraic condition {\bf A1}, 
which is equivalent to the Cauchy property.
Condition {\bf A1} is the condition that the matrix $\hat P^{d}: 
V_{\lambda}\rightarrow  (\Complexes^{d}\otimes V_{\lambda})_{\Delta}$
be injective.  Since we picked out the direction
$d$, $\hat P^{d}$ is not fully $\mathbf{so}(d)$-invariant, but 
it is $\mathbf{so}(d-1)$-invariant.
The idea is thus to decompose all $\mathbf{so}(d)$ representations
into $\mathbf{so}(d-1)$ representations.

We can obtain necessary conditions for $\hat P^d$ to be injective
by  using Schur's Lemma.
Any non-trivial irreducible
$\so(d-1)$ representation $V_\mu$ that occurs in $V_\lambda$
must also occur in $(\Complexes^{d}\otimes V_{\lambda})_{\Delta}$.
Otherwise the $\so(d-1)$ invariant map $\hat P^d$ must map it to the trivial representation,
which means that $\hat P^d$ cannot be injective.
Therefore,
a necessary condition for $\hat 
P^{d}$ to be injective is
\begin{itemize}
\item[\COne] Every inequivalent irreducible 
$\mathbf{so}(d-1)$-representation $V_\mu$ that occurs in $V_{\lambda}$ must 
also occur in $(\Complexes^{d}\otimes V_{\lambda})_{\Delta}$.
\end{itemize}
If we demand that the representation $V_\lambda$ lead to
a unitary representation of the full conformal group,
it is necessary that $\Delta$ is the largest eigenvalue
of $\hat M$.
\begin{itemize}
\item[\CTwo] For $(V_\lambda,\,\Delta)$ to be a unitary
conformal representation, $\Delta$ must be the largest eigenvalue
of $\hat M$.
\end{itemize}

Let us first discuss condition \COne. 
For $\lambda$ non-trivial, the branching rules given in
(\ref{eq:branching2}) imply that
the only $\lambda'$ which satisfy this necessary 
condition are
\eq
\label{eq:lambdaprimelist}
\begin{array}{lrl}
d=2n:
\\
& \mathbf{A_{2n}}: & \lambda'=\lambda+\epsilon_{1}
\\[1ex]
& \mathbf{B_{2n}^{+}}: & \lambda' = \lambda - \epsilon_{n}\,,\quad
\lambda_{n}>0\,,
\\[1ex]
& \mathbf{B_{2n}^{-}}: &
\lambda' = \lambda + \epsilon_{n}\,,\quad
\lambda_{n}<0\,,
\\[2ex]
d=2n+1:
\\
& \mathbf{A_{2n+1}}: &\lambda'=\lambda+\epsilon_{1}
\\[1ex]
& \mathbf{C_{2n+1}}: & \lambda'=\lambda\,,
\end{array}
\en
To see this, note that for instance if $\lambda'=\lambda- \epsilon_j$, $j<n$,
then $\mu=\lambda$ is not in $\lambda'$; for $\lambda'=\lambda+\epsilon_j$,
$\mu=(\lambda_1,\ldots,\lambda_j,\lambda_j,\ldots)$ is in $\lambda$
but not in $\lambda'$. Similar arguments eliminate the
other cases.   
For $\lambda=0$, the only $\lambda'$ is $\Complexes^{d}$, 
which satisfies the necessary condition.

\subsection{Assuming unitarity}\label{ss:classifyUnitary}
Let us now show that the unitary Cauchy fields with spin $V_{s}$ 
listed in section \ref{s:summaryCauchyfields}
are the only unitary Cauchy fields that satsify condition \CTwo.
This completes the classification of unitary Cauchy fields.

First
note from (\ref{evenpos}) -- (\ref{oddzero}) that
unless $\lambda=0$, $\lambda'=\lambda+\epsilon_1$ never
leads to the largest eigenvalue. In the following always
assume that $\lambda \neq 0$.

Next, for $d=2n$ consider the case $\lambda_{n}>0$, $\lambda'=\lambda-\epsilon_n$.
From (\ref{evenpos}) we see that $\lambda'$ can only
be the largest eigenvalue if none of the $\lambda-\epsilon_j$
are representations of $so(d)$. This is only 
the case if $\lambda = (|s|,\ldots,|s|,|s|)$.
Similarly, for $\lambda_{n}<0$, $\lambda'=\lambda+\epsilon_n$,
we see from (\ref{evenneg}) that  $\lambda'=\lambda+\epsilon_n$ is only
the largest eigenvalue if $\lambda = (|s|,\ldots,|s|,-|s|)$.
This establishes the claim for even $d$.

For odd $d$, $\lambda'=\lambda$ has to lead to
the largest eigenvalue. From (\ref{oddpos}) no
representations with  $\lambda-\epsilon_j$
can appear, so that $\lambda = (|s|,\ldots,|s|,|s|)$.
Moreover $\lambda-\epsilon_n$ must not appear
either, which is only the case if $s = 1/2$.
This establishes the claim for odd $d$.

\subsection{The non-unitary cases: $d=3$ and $d=4$}
Let us now investigate the injectivity condition
\COne if we do not require unitarity. We did not
develop the general theory, but the special cases
$d=3$ and $d=4$ are relatively straightforward to work out.

For $d=3$, the problem reduces to decomposing $\so(3)$ into
$\so(2)$ representations, which is decomposing $su(2)$
into $\u(1)$ representations. Since the latter are 1 dimensional
it is enough to check that $P^{d}_{\lambda,\lambda'}$
is non-vanishing on each $\u(1)$ representation $\mu$ that occurs
in $\lambda$,
which is equivalent to the non-vanishing of the relevant
Clebsch-Gordan coefficient between the fundamental $\C^3$ of $so(3)$, 
$\lambda$, and $\lambda'$.
In terms of Wigner 3-$j$ symbols
the condition is thus
\eq
\begin{pmatrix}
1 & \lambda & \lambda' \\
0 & \mu & -\mu
\end{pmatrix}
\ne 0\,, \qquad
|\mu| \le \lambda\ ,
\;\; \lambda-\mu\in \Integers
\, ,
\en
where by assumption $\lambda'$  occurs in the fusion of
$\lambda$ with $\C^3$. There is then just one additional selection rule
on the 3-$j$ symbols, namely
\eq
\begin{pmatrix}
1 & \lambda & \lambda' \\
0 & \mu & -\mu
\end{pmatrix}
=0
\quad \text{iff}
\quad
1+\lambda+\lambda' \in 2\Integers+1
\text{ and }
\mu = 0\,.
\en
This never affects the case $\mathbf{A_{3}}$,
but if $\lambda=\lambda'$ is integer, then $P^{d}_{\lambda,\lambda'}$
vanishes on $\mu=0$.
Condition \COne is thus satisfied for
\eq
\begin{array}{lrll}
d=3
\\
& \mathbf{A_{3}}: &\lambda'=\lambda+\epsilon_{1}
& \text{injective}
\\[1ex]
& \mathbf{C_{3}}: & \lambda'=\lambda
& \text{injective iff } \lambda\in \frac12 + \Integers
\,.
\end{array}
\en

For $d=4$, we show that the map $P^{d}_{\lambda,\lambda'}$ is injective
in all three of the non-unitary cases 
$\mathbf{A_{4}}$, $\mathbf{B_{4}^{+}}$, $\mathbf{B_{4}^{-}}$.

Use $\mathbf{so}(4) = \mathbf{sl}(2) \oplus \mathbf{sl}(2) $.
The representation $(\lambda_{1},\lambda_{2})$ of $\mathbf{so}(4)$ is 
the representation $(j)\otimes (k)$ of $\mathbf{sl}(2) \oplus 
\mathbf{sl}(2) $, with
\eq
\lambda_{1} = j+k\,, \qquad
\lambda_{2} = j-k\,.
\en
The fundamental representation $(\frac12)$ of $\mathbf{sl}(2)$ is 
$\Complexes^{2}$ with invariant anti-symmetric tensor $\epsilon^{ab}$, $a,b=1,2$.
Take as basis for the $\mathbf{sl}(2)$ representation $(j)$ the 
symmetric tensors of rank $2j$ on $\Complexes^{2}$.  
For the second $\sl2$, take $(\Complexes^{2})^{*}$ as the 
fundamental representation.
This gives a 
basis for $\lambda$ as tensors on $\Complexes^{2}$,
\eq
t^{a_{1}\cdots a_{2j}}_{b_{1}\cdots b_{2k}}
\en
symmetric separately in the $a_{i}$ and in the $b_{i}$.
The representation $\Complexes^{d}$ has basis $t^{a}_{b}$.
The $\mathbf{so}(3)$ invariant is $\delta^{a}_{b}$.

The case $\mathbf{A_{4}}$ is $j' = j+\frac12$, \;\;$k'=k+1/2$.  
The projection $\hat P$ is thus given by the map
\be
\hat P : \Complexes^d \otimes V_\lambda \rightarrow (\Complexes^d \otimes V_\lambda)_{\lambda'}\ ,
\qquad (t^a_b, t^{a_{1}\cdots a_{2j}}_{b_{1}\cdots b_{2k}}) \mapsto \mathbf{Sym}_{a} \mathbf{Sym}_{b} 
\;
\left ( t^{a_{1}\cdots a_{2j}}_{b_{1}\cdots b_{2k}}
t^{a}_{b}
\right )
\ee
The 
map $P^{d}_{\lambda,\lambda'}$ is
\eq
t^{a_{1}\cdots a_{2j}}_{b_{1}\cdots b_{2k}}
\mapsto
\mathbf{Sym}_{a} \mathbf{Sym}_{b} 
\;
\left ( t^{a_{1}\cdots a_{2j}}_{b_{1}\cdots b_{2k}}
\delta^{a_{2j+1}}_{b_{2k+1}}
\right )
\en
which is pretty clearly injective.  To see this, note that the 
irreducible components under $\mathbf{so}(d-1)$ have basis
\eq
t^{a_{1}\cdots a_{2j}}_{b_{1}\cdots b_{2k}}
= \mathbf{Sym}_{a} \mathbf{Sym}_{b} 
\left ( w^{a_{1}\cdots a_{2j-r}}_{b_{1}\cdots b_{2k-r}} 
\delta^{a_{2j-r+1}}_{b_{2k-r+1}} \cdots \delta^{a_{2j}}_{b_{2k}}
\right )
\,,\qquad
 w^{a_{1}\cdots a_{2j-r-1} c}_{b_{1}\cdots b_{2k-r-1}c}  =0
\,.
\label{eq:sod-1basis}
\en
In this basis, $P^{d}_{\lambda,\lambda'}$ is just the identity on 
each $\mathbf{so}(d-1)$ component of $\lambda$.

The case $\mathbf{B_{2n}^{+}}$ is
$j> k>0$,\;\;
$j' = j-\frac12$, \;\;$k'=k+1/2$.
The 
map $P^{d}_{\lambda,\lambda'}$ is
\eq
t^{a_{1}\cdots a_{2j}}_{b_{1}\cdots b_{2k}}
\mapsto
\mathbf{Sym}_{b} 
\;
\left (t^{a_{1}\cdots a_{2j}}_{b_{1}\cdots 
b_{2k}}\epsilon_{a_{2j}b_{2k+1}}
\right )
\,.
\en
Again look at the action of $P^{d}_{\lambda,\lambda'}$ on
the $\mathbf{so}(d-1)$ component of $\lambda$ with 
basis elements given in (\ref{eq:sod-1basis}),
\eq
t^{a_{1}\cdots a_{2j}}_{b_{1}\cdots b_{2k}}
\mapsto
\mathbf{Sym}_{a} \mathbf{Sym}_{b} 
\left ( w^{a_{1}\cdots a_{2j-r-1}a_{2j}}_{b_{1}\cdots b_{2k-r}} 
\epsilon_{a_{2j} b_{2k-r+1}}
\delta^{a_{2j-r}}_{b_{2k-r+2}} \cdots \delta^{a_{2j-1}}_{b_{2k+1}}
\right )
\,.
\en
We only have to show that this is non-zero.  Take the tensor $w$ to 
have only one non-zero component, $w^{11\cdots}_{2 2\cdots}=1$.
Then $w^{11\cdots}_{2 2\cdots}\epsilon_{12}\ne 0$.
So $P^{d}_{\lambda,\lambda'}$ is injective.

The case $\mathbf{B_{4}^{-}}$ is the same
as $\mathbf{B_{4}^{+}}$ with $j\leftrightarrow k$.

So, for $d=4$, in all three of the non-unitary cases 
$\mathbf{A_{4}}$, $\mathbf{B_{4}^{+}}$, $\mathbf{B_{4}^{-}}$,
the map $P^{d}_{\lambda,\lambda'}$ is injective.

\subsection{The non-unitary case: $d>4$}
\label{s:nonunitarydabove4}

The question is, for $V_{\lambda}$ an irreducible representation of 
$\so(d)$, and $V_{\lambda'}$ one of the irreducible components of 
$\C^{d}\otimes V_{\lambda}$ listed in (\ref{eq:lambdaprimelist}),
is the map $P^{d}_{\lambda,\lambda'}:V_{\lambda}\rightarrow 
V_{\lambda'}$ injective, where
\eq
P^{d}_{\lambda,\lambda'}(v) = P_{\lambda,\lambda'}(\hat 
e_{d}\otimes v)
\en
where
\eq
P_{\lambda,\lambda'} : \C^{d}\otimes V_{\lambda} 
\rightarrow V_{\lambda'}
\en
is the projection on the component $V_{\lambda'}$.
We cannot answer the question, but we convey a suggestion from 
N.~Wallach:
\begin{quote}
In the case of $\lambda+\epsilon_1$ mapping into $\lambda \otimes
\epsilon_1$, $P^{d}_{\lambda,\lambda'}$ is injective since it is just Cartan
multiplication (which is multiplication in an integral domain).  
In the even dimensional case $\mathbf{B_{2n}^{\pm}}$,  $P^{d}_{\lambda,\lambda'}$ is
adjoint to Cartan multiplication (which is multiplication in an
integral domain).  That is we look at $\lambda$ mapping into
$(\lambda+\epsilon_1)\otimes \epsilon_1$, by realizing
$\lambda+\epsilon_1$ in $\lambda \otimes \epsilon_1$ and contracting. 
\cite{WallachPrivate}
\end{quote}


\section{First-order over-determination conditions for $V=V_{s}$}
\label{app:overdet}

As an illustration, let us now work out the first-order 
over-determination
conditions explicitly for the Cauchy representations $V_s$.
Take $d=2n$ and $V=V_{s}$, $s\ne 0$.  

Let us first work out the
eigenvalues of $P^d_d$ and $\overleftrightarrow{P}$.
Since they are self-adjoint,
we can decompose
$\Complexes\otimes V$  into eigenspaces of
$\hat P^{d}_{d}$
and decompose
$\Complexes^{d-1}\otimes V$ into eigenspaces of
$\overleftrightarrow{P}$
\ateq{2}{
	\Complexes\otimes V &= \mathop\oplus_{\lambda} V_{\lambda} 
	\,,\qquad & \left.\hat P^{d}_{d}\right|_{V_{\lambda}}&=\lambda\\
	\Complexes^{d-1}\otimes V &= \mathop\oplus_{\lambda} W_{\lambda}
	\,,\qquad & \left.\overleftrightarrow{P}\right|_{W_{\lambda}}&=\lambda
	\, ,
}
but since they are not projectors, they can have eigenvalues different from 
$\lambda=0,1$.
Equations (\ref{eq:Phatproj4})--(\ref{eq:Phatproj1}) become,
for $v_{\lambda}\in V_{\lambda}$ and $w_{\lambda}\in W_{\lambda}$,
\aeq{
	\vec P^{\dagger} \vec P v_{\lambda} &= \lambda(1-\lambda) v_{\lambda}
	\label{eq:Phatprojlambda1}\\
	(\hat P^{d}_{d}  + \lambda -1) \vec P^{\dagger} w_{\lambda} &= 0
	\label{eq:Phatprojlambda2}\\
	(\overleftrightarrow{P}+ \lambda -1) \vec P v_{\lambda} &= 0
	\label{eq:Phatprojlambda3}\\
	\vec P \vec P^{\dagger} w_{\lambda}  &= \lambda(1-\lambda) w_{\lambda}
	\label{eq:Phatprojlambda4}
	\,.
}
The Cauchy condition $\Ker \hat P_d^d = 0$ means that $V_0 = 0$,
so that (\ref{eq:Phatprojlambda1}) and (\ref{eq:Phatprojlambda4})
respectively imply that
\ateq{2}{
	\Ker\,\vec P &= V_{1}
	\,,\qquad&
	\Im\,\vec P^{\dagger} &= \sum_{\lambda\ne 0,1}V_{\lambda}
	\\
	\Ker\,\vec P^{\dagger} &= W_{0}\oplus W_{1}
	\,,\qquad&
	\Im\,\vec P  &= \sum_{\lambda\ne 0,1}W_{\lambda}
}
whereas the second and third equation are equivalent to
\eq\label{WandV}
\vec P V_{\lambda} = W_{1-\lambda}
\,,\qquad
\vec P^{\dagger} W_{\lambda} = V_{1-\lambda}
\,,\qquad
\lambda\ne 0,1
\,.
\en
Let us now evaluate these expression for Cauchy representations
$V_s$. The projection $\hat P$ is
\eq
\hat P = \frac{\hat M -  (n-1-\Delta)}{\Delta -  (n-1-\Delta)}
\,,
\en
with $\Delta = n-1+|s|$ in the unitary case and $\Delta = -|s|$ in 
the non-unitary case. 
Since $\hat M^{d}_{d}=0$, $\hat P^{d}_{d}$ has only one eigenvalue,
\eq
\hat P^{d}_{d} = \frac{ -  (n-1-\Delta)}{\Delta -  (n-1-\Delta)}
\,.
\en
From (\ref{WandV})
it follows that $\overleftrightarrow{P}$ has at most eigenvalues
\eq
\lambda = 1 - \frac{ -  (n-1-\Delta)}{\Delta -  (n-1-\Delta)} =
\frac{\Delta}{2\Delta -  (n-1)}
\en
and $\lambda= 0$ and 1.
Having established that, let us work out the eigenspaces
of $\overleftrightarrow{P}$ explicitly. Using
\eq
\overleftrightarrow{P} = \frac{\hat M^{d-1} -  (n-1-\Delta)}{\Delta -  (n-1-\Delta)}
\,,
\en
the strategy is of course to use the $\so(d-1)$
invariance of $\hat M^{d-1}$ to express it
in terms of $\so(d-1)$ Casimirs. Decomposing
$V_s$ into representations of $\so(d-1)$,
\eq
V_{s} = V_{\mu_{|s|}}
\,,\qquad
\mu_{|s|} = (|s|,\ldots, |s|)
\en
\eq
\Complexes^{d-1}\otimes V_{s} =
V_{\mu_{|s|} } \oplus
V_{\mu_{|s|}+\epsilon_{1} } \oplus
V_{\mu_{|s|}-\epsilon_{n-1} }
\en
where the right-most summand does not occur for $|s|=1/2$.
The quadratic Casimirs are
\aeq{
C_{2}^{d-1}(V_{\mu}) &= \frac1{2}\sum_{i=1}^{n-1}  \mu_{i} 
(\mu_{i}+d-1-2i)
\\
C_{2}^{d-1}(\Complexes^{d-1}) &= \frac12(d-2) = n -1
\\
C_{2}^{d-1}(V_{\mu_{|s|} }) &=
\frac12 (n-1) |s|(|s|+n-1)
\\
C_{2}^{d-1}(V_{\mu_{|s|}+\epsilon_{1} }) &=C_{2}^{d-1}(V_{\mu_{|s|} })
+ |s|+n-1
\\
C_{2}^{d-1}(V_{\mu_{|s|}-\epsilon_{n-1} }) &=C_{2}^{d-1}(V_{\mu_{|s|} })
-|s|
}
Writing $\hat M^{d-1}$ as
\eq
\hat M^{d-1}
= \mathbf1\otimes C_{2}^{d-1}(V_{\mu_{|s|} }) + 
C_{2}^{d-1}(\Complexes^{d-1})  \otimes \mathbf1
- C_{2}^{d-1}(\Complexes^{d-1} \otimes V_{\mu_{|s|} })
\,,
\en
its eigenspaces and eigenvalues are 
\ateq{2}{
& n-1
&\qquad &\text{on }V_{\mu_{|s|} }
\\
&  -|s|
&\qquad &\text{on }V_{\mu_{|s|}+\epsilon_{1} }
\\
&n-1  + |s|
&\qquad &\text{on }V_{\mu_{|s|}-\epsilon_{n-1} }\ ,
}
so that the eigenvalues of $\overleftrightarrow{P}$ are
\ateq{2}{
&  \frac{\Delta}{2\Delta-(n-1)}
&\qquad &\text{on }V_{\mu_{|s|} }
\\
& \frac{\Delta-|s|-n+1}{2\Delta-(n-1)}
&\qquad &\text{on }V_{\mu_{|s|}+\epsilon_{1} }
\\
&\frac{\Delta+|s|}{2\Delta-(n-1)}
&\qquad &\text{on }V_{\mu_{|s|}-\epsilon_{n-1} }
}
For the unitary case, $\Delta = |s|+n-1$,
these eigenvalues of $\overleftrightarrow{P}$ are
\ateq{2}{
& \frac{\Delta}{2\Delta-(n-1)}
&\qquad &\text{on }V_{\mu_{|s|} }
\\
& 0
&\qquad &\text{on }V_{\mu_{|s|}+\epsilon_{1} }
\\
&1
&\qquad &\text{on }V_{\mu_{|s|}-\epsilon_{n-1} }
}
For the non-unitary case, $\Delta = -|s|$,
\ateq{2}{
&  \frac{\Delta}{2\Delta-(n-1)}
&\qquad &\text{on }V_{\mu_{|s|} }
\\
& 1
&\qquad &\text{on }V_{\mu_{|s|}+\epsilon_{1} }
\\
&0
&\qquad &\text{on }V_{\mu_{|s|}-\epsilon_{n-1} }
}
So, for the unitary case, (\ref{eq:overdetermine}) leads to an 
over-determination condition when 
$|s|>1/2$,
\eq
\mathit{Proj}_{V_{\mu_{|s|}-\epsilon_{n-1} }}(\vec\partial\phi) = 0
\,.
\en
For the non-unitary case, there is an over-determination condition for 
all $|s|$,
\eq
\mathit{Proj}_{V_{\mu_{|s|}+\epsilon_{1} }}(\vec\partial\phi) = 0
\,.
\en
In both cases, the over-determination condition can be written
\eq
\overleftrightarrow{P} 
\left(\overleftrightarrow{P}-\frac{\Delta}{2\Delta-(n-1)}\right)
\vec\partial\phi = 0
\,,
\en
or, equivalently
\eq
\left(\hat M^{d-1} -(n-1)\right)\left(\hat M^{d-1} -(n-1)+\Delta\right)
\vec\partial\phi = 0
\,.
\en

\bibliographystyle{../ytphys}
\bibliography{../ref}

\end{document}